\documentclass[pre,aps,twocolumn,showpacs,reprint,floatfix,superscriptaddress,nofootinbib]{revtex4-1}

\usepackage{graphicx}
\usepackage{amssymb,amsfonts,amsmath}
\usepackage{color}
\usepackage[normalem]{ulem}
\usepackage{subfigure}
\usepackage{comment}

\newcommand{\lrangle}[1]{\langle{#1}\rangle}

\newcommand{\vinf}{v_\infty}

\usepackage{hyperref}

\begin{document}
\title{Hallmarks of the Kardar-Parisi-Zhang universality 
class elicited by scanning probe microscopy}
\author{Sidiney G. Alves}
\affiliation{Departamento de F\'isica e Matem\'atica, Universidade Federal de S\~ao Jo\~ao Del-Rei,
36420-000, Ouro Branco, Minas Gerais, Brazil}
\author{Clodoaldo I. L. de Araujo}
\author{Silvio C. Ferreira}
\affiliation{Departamento de F\'{\i}sica, Universidade Federal de Vi\c{c}osa,
36570-000, Vi\c{c}osa, Minas Gerais, Brazil}


\begin{abstract}
Scanning probe microscopy is a fundamental technique for the analysis of
surfaces. In the present work, the interface statistics of surfaces scanned with
a probe tip is analyzed for both \textit{in silico} and experimental systems
that, in principle, \textit{do not} belong to the prominent Kardar-Parisi-Zhang
(KPZ) universality class. We observe that some features such as height, local
roughness and extremal height distributions of scanned surfaces quantitatively
agree with the KPZ class with good accuracy. The underlying mechanism behind
this artifactual KPZ class is the finite size of the probe tip, which does not
permit a full resolution of neither deep valleys nor sloping borders of
plateaus. The net result is a scanned profile laterally thicker and higher than
the original one implying an excess growth, a major characteristic of the KPZ
universality class. Our results are of relevance whenever  either the normal or
lateral characteristic lengths of the surface are comparable with those of the
probe tip. Thus our finds can be relevant, for example, in experiments where
sufficiently long growth times cannot be achieved or  in mounded surfaces with
high aspect ratio.
\end{abstract}

\flushbottom
\maketitle

\thispagestyle{empty}

\section{Introduction}

Universality beyond scale invariance and critical exponents is well
established in both equilibrium~\cite{privman1990finite,binder1981finite}
and nonequilibrium~\cite{henkel2008non} critical systems. Universal 
fluctuations underlying the interface growth were sown by Krug \textit{et
al}.~\cite{KrugPRA92} in the framework of the Kardar-Parisi-Zhang (KPZ)
universality class~\cite{KPZ}, represented by the non-linear stochastic
equation governing the surface height evolution $h(\mathbf{x},t)$
\begin{equation}
 \frac{\partial h}{\partial t} = \nu \nabla^2 h +\frac{\lambda}{2}|\nabla h|^2+\eta.
 \label{eq:kpz}
\end{equation}
The first and third terms in the right-hand side of Eq.~\eqref{eq:kpz}
represent, respectively, the surface tension and a white noise given by
$\lrangle{\eta}=0$ and $\lrangle{\eta(\mathbf{x},t)\eta(\mathbf{x}',t')} = 2D
\delta(\mathbf{x}-\mathbf{x}') \delta(t-t')$. The second term represents
a fundamental benchmark of the KPZ class~\cite{KPZ}: the surface grows {locally}
faster (or slower if $\lambda<0$) than the rate of particles arriving at
the surface, resulting in an \textit{excess velocity}. The KPZ equation is
a fundamental example that follows the Family-Vicsek scaling~\cite{Family85}, in
which the variance of the height profile (also called squared interface
width or roughness) obeys the ansatz $w^2(l,t) = t^{2\beta}
f(l/t^{\beta/\alpha})$ as a function of time $t$ and scale of measurement
$l$. The function $f(x)$ scales as $f(x)\sim x^{2\alpha}$ for $x\ll 1$ and
$f(x)\sim \mbox{const}$ if $x\gg1$. The growth ($\beta$) and roughness
($\alpha$) exponents yield the interface scaling in  the transient and
stationary regimes  as $w\sim t^\beta$ and $w\sim l^\alpha$ for $t\ll
l^{\alpha/\beta}$ and $t\gg l^{\alpha/\beta}$, respectively~
\cite{barabasi}.

{Consider a $d_s+1$-dimensional system  where $d_s$ is the dimension 
of the growth substrate.}
Krug \textit{et al}.~\cite{KrugPRA92} proposed that a flat initial
condition in 1+1-dimensional KPZ surfaces asymptotically leads to the
ansatz
\begin{equation}
h=\vinf t + \mbox{sgn}(\lambda) (\Gamma t)^\beta \chi,
 \label{eq:kpz:ansatz}
\end{equation}
where $\vinf$ is the asymptotic growth velocity, $\Gamma$  is a parameter
associated to the amplitude of the interface width, $\beta=1/3$ is the growth
exponent in $d=1+1$, and $\chi$ is a universal stochastic quantity whose
distribution does not depend on the microscopic details of the model but does on
the boundary and initial conditions~\cite{PraSpo1}. It is worth to remark that
analytical~\cite{Johansson-CMP2000,SasaSpo1} and experimental~\cite{TakeuchiPRL}
confirmations of the KPZ ansatz were obtained first in curved geometry.
Posteriorly, this conjecture was confirmed in several independent
works~\cite{Calabrese,Oliveira12,TakeuchiSP}. The distribution of $\chi$ is
known in $d=1+1$ for a flat initial condition~\cite{Calabrese} as the
Tracy-Widom distribution of the largest eigenvalue of  the Gaussian orthogonal
ensemble (GOE)~\cite{TW}.

Despite of the lack of rigorous results in $d=2+1$, a meticulous analysis of
several models~\cite{MF12,Oliveira13,MF13}, accepted as belonging to the
KPZ class, and the numerical integration of the KPZ equation
itself~\cite{MF12} strongly indicate that the ansatz given by
Eq.~\eqref{eq:kpz:ansatz} also holds for the three-dimensional case with the
corresponding growth exponent $\beta=0.241$ and a new universal stochastic
quantity $\chi$ with mean $\lrangle{\chi}\simeq - 0.83$, variance
$\lrangle{\chi^2}_c\simeq 0.237$, skewness
$S=\lrangle{\chi^3}_c/\lrangle{\chi^2}_c^{1.5}\simeq0.43$, and kurtosis
$K=\lrangle{\chi^4}_c/\lrangle{\chi^2}_c^2\simeq 0.35$~\cite{
MF12,Oliveira13,Alves12}. Hereafter, $\lrangle{X^n}_c$ is the $n$th cumulant
of $X$. {Very recently, an analogous of the KPZ ansatz for a universality class
different from KPZ has been proposed~\cite{Carrasco2016b}.}

Earlier attempts of experimental realizations of the $2+1$-KPZ class through
scaling exponents were not fully conclusive due to experimental limitations as,
for example, short times and length scales  as well as instabilities caused by
interlayer diffusion barriers~\cite{evans2006} (See also
Refs.~\cite{michely2004islands,meakin1998fractals,Krim95} for more examples and
references) and only a few experiments reporting scaling exponents consistent
with the 2+1-KPZ class  are known~\cite{Palasantzas02,Ojeda2000}. In particular,
the deposition of organic films on silicon substrates~\cite{Palasantzas02}
provided $\beta=0.28(4)$ and $\alpha=0.45(5)$ (using approximately 1 decade in
power law regressions), which are consistent, within uncertainties, with the
accepted $2+1$-KPZ exponents $\beta\approx 0.24$ and $\alpha\approx
0.39$~\cite{Kelling}. The results of Ref.~\cite{Palasantzas02} were recently
revisited~\cite{MF2014} and height distributions shifted to null mean and
unitary variance are fitted by the distributions of the $2+1$-KPZ class within
four orders of magnitude. Also, preceding the studies of~\cite{MF2014}, rescaled
KPZ height distributions were found in the growth of CdTe semiconductor
films~\cite{almeida2013,Almeida2015}. The experimental realization of  the
$2+1$-KPZ class was further supported in \cite{almeida2013,Almeida2015,MF2014}
by the agreement between local roughness~\cite{Racz} and extremal
height~\cite{Raychaudhuri,Majumdar04} distributions with the corresponding
$2+1$-KPZ curves. Furthermore, spatial correlators in good agreement with
2+1-KPZ class were also presented~\cite{MF2014}. Even in cases where the scaling
exponents deviate considerably from the KPZ ones, rescaled distributions have
been pointed as robust and little sensitive to finite time and size
effects~\cite{Almeida2015}.

A common feature in Refs.~\cite{Palasantzas02,almeida2013,Almeida2015,MF2014} is
that the analyzed surfaces were acquired through atomic force microscopic (AFM)
technique. However, the nonlinear smoothing effects caused by the finite size of
the probe tip can introduce subtle but relevant biases in the experimental
determination of scaling exponents of self-affine
surfaces~\cite{Lechenault,villarrubia94}. Motivated by the experimental
observations of KPZ distributions and the well known smoothing effect inherent
to the scanning probe microscopy (SPM) technique~\cite{Lechenault}, we
investigated the role played by a probe tip in the statistics of surfaces. We
analyzed computer models and the electrodeposition of iron on silicon
substrates, systems that do not belong to the KPZ universality class. We observe
that the scaled KPZ distributions can be found in the scanned surfaces when
the original ones \textit{do not belong} to the KPZ class.  The growth exponents
become closer to the KPZ value for scanned than for original interfaces. Also,
the average growth velocity of the scanned surface is apparently consistent with
the KPZ ansatz. We observed the aforementioned accordance when the scanning
probing tip scales are comparable with the characteristic lengths of the surface
and introduce a non-negligible excess of height in the profiles.

The paper layout is organized as follows.  In section~\ref{sec:spm_surface}, we
discuss the effects of a scanning probe tip on the statistics of growing
surfaces using models which are known as not belonging to the KPZ class. In
section~\ref{sec:eletro}, we analyze the electrodeposition of iron on silicon
substrates as an experimental example of KPZ distributions produced by tip
effects. In section~\ref{sec:discuss}, some strategies to rule out false
positives for KPZ are discussed and in section~\ref{sec:conclusions} we
summarize  our conclusions and draw some prospects.

\section{SPM surfaces and the KPZ class}
\label{sec:spm_surface}

\begin{figure}[!hb]
 \centering
 \includegraphics[width=3.5cm]{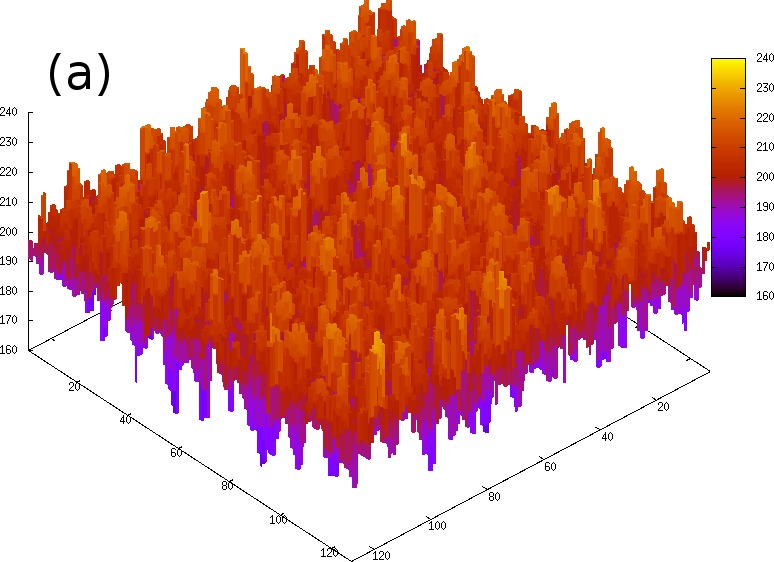}   \includegraphics[width=3.5cm]{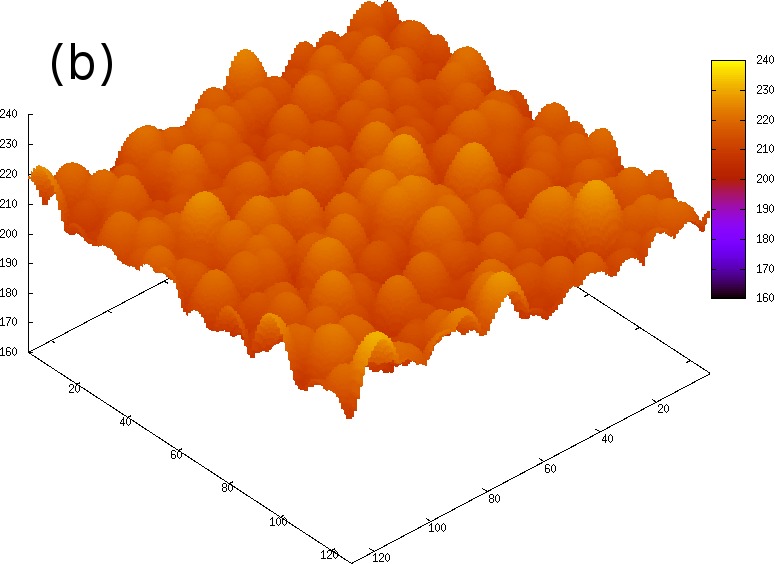} 
 \\ \medskip 
 \includegraphics[width=8.0cm]{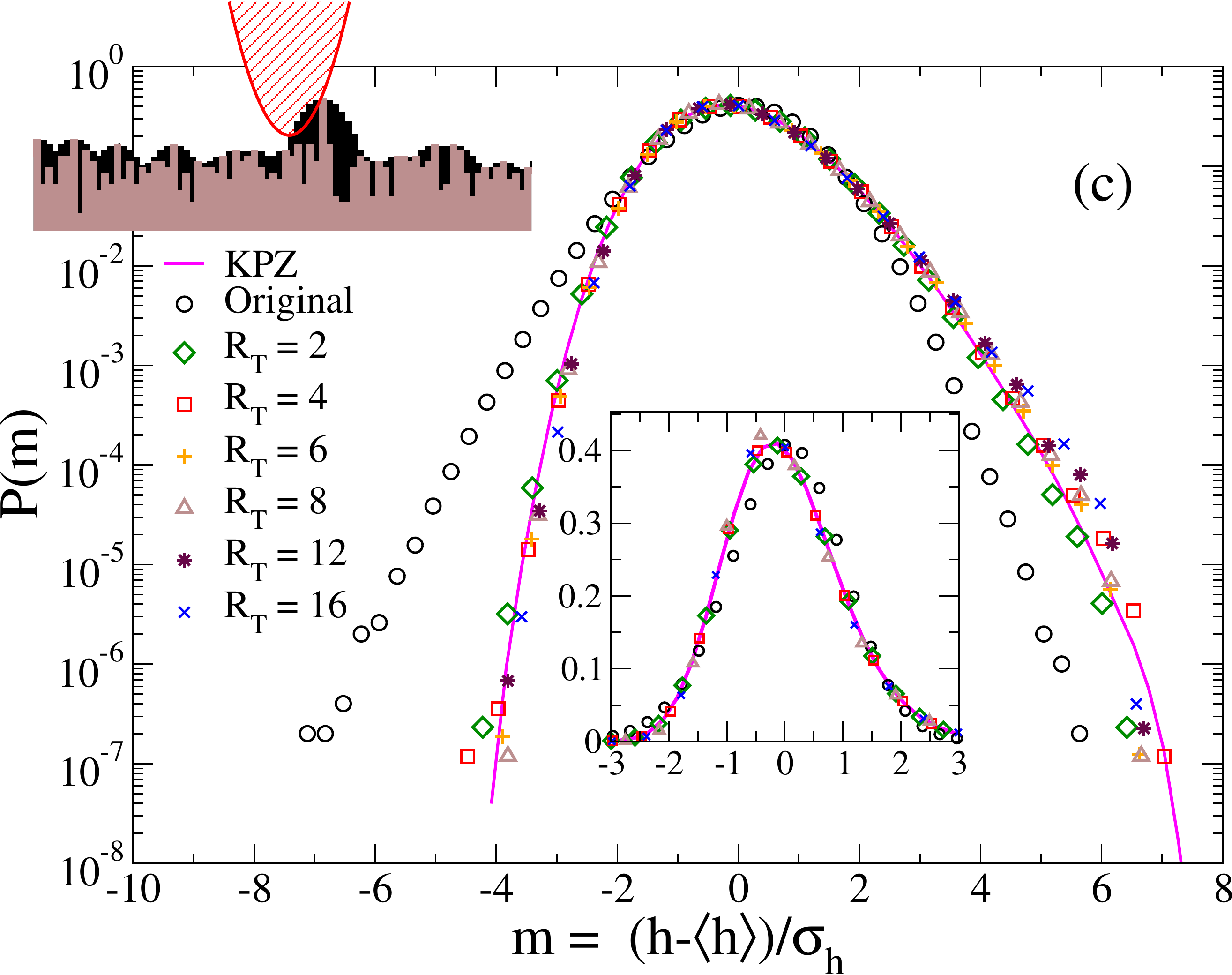} 
 % ph_dt1d.pdf: 758x561 pixel, 72dpi, 26.74x19.79 cm, bb=0 0 758 561
 \caption{ Original (a) and scanned (b)
 surfaces generated with the $2+1$-MBE model. (c) Height distributions rescaled
 to mean zero and variance 1 for parabolic tips of different sizes. The system
 size is $2048\times2048$ and the growth time $t=250$. A $2+1$-KPZ (RSOS model)
 distribution  is also shown. Inset: the same distributions in linear scales.
 Upper insertion illustrates the effect of the scanning probe tip.}
\label{fig:phMBE2d}
\end{figure}

The rationale for a potential false positive of KPZ class is
pictorially shown in the upper insertion of Fig.~\ref{fig:phMBE2d}(c)
where the scanning of an originally rough surface brings forth a
smoothed profile; See also Figs. 1 and 2 in Ref.~\cite{Lechenault}.
Here, more important than the short wavelength smoothing, which
affects the determination of the roughness exponent~\cite{Lechenault},
is that the profiles are laterally thicker having a local height
higher than the original one and it represents a central mechanisms
behind the KPZ equation: normal growth and excess
velocity~\cite{KPZ,barabasi}. Might this artifact introduce traits of
the KPZ class? To shed light in this issue we analyzed surfaces
numerically obtained with a simple irreversible model for molecular
beam epitaxy (MBE), in which  each deposited particle starts to
diffuse randomly  with probability $p^{n+1}$,  where $n$ is the number
of {bonds in the same deposition layer}, or gets
irreversibly stuck with the complementary probability. A step
barrier~\cite{evans2006} for diffusion is also included as a
probability $P_D$ to diffuse downwardly leading to  mounded instead of
self-affine surfaces, as shown in Fig.~\ref{fig:phMBE2d}(a). We
present results for $p=P_D=0.9$. We also analyzed the classic
Wolf-Villain (WV)~\cite{WV} and Das Sarma-Tamborenea (DT)~\cite{DT}
models (see also~\cite{barabasi} for model definitions), which are
known as not belonging to the KPZ class. In $d=2+1$ dimensions, these
models have upward currents {induced
by the configurations of the neighborhood of particles
deposited on steps} and asymptotically produce mounded
surfaces~\cite{Chatraphorn}. All simulations were done on square
lattices with
periodic boundary conditions.

\begin{figure}[ht]
 \centering
 \includegraphics[height=7.5cm]{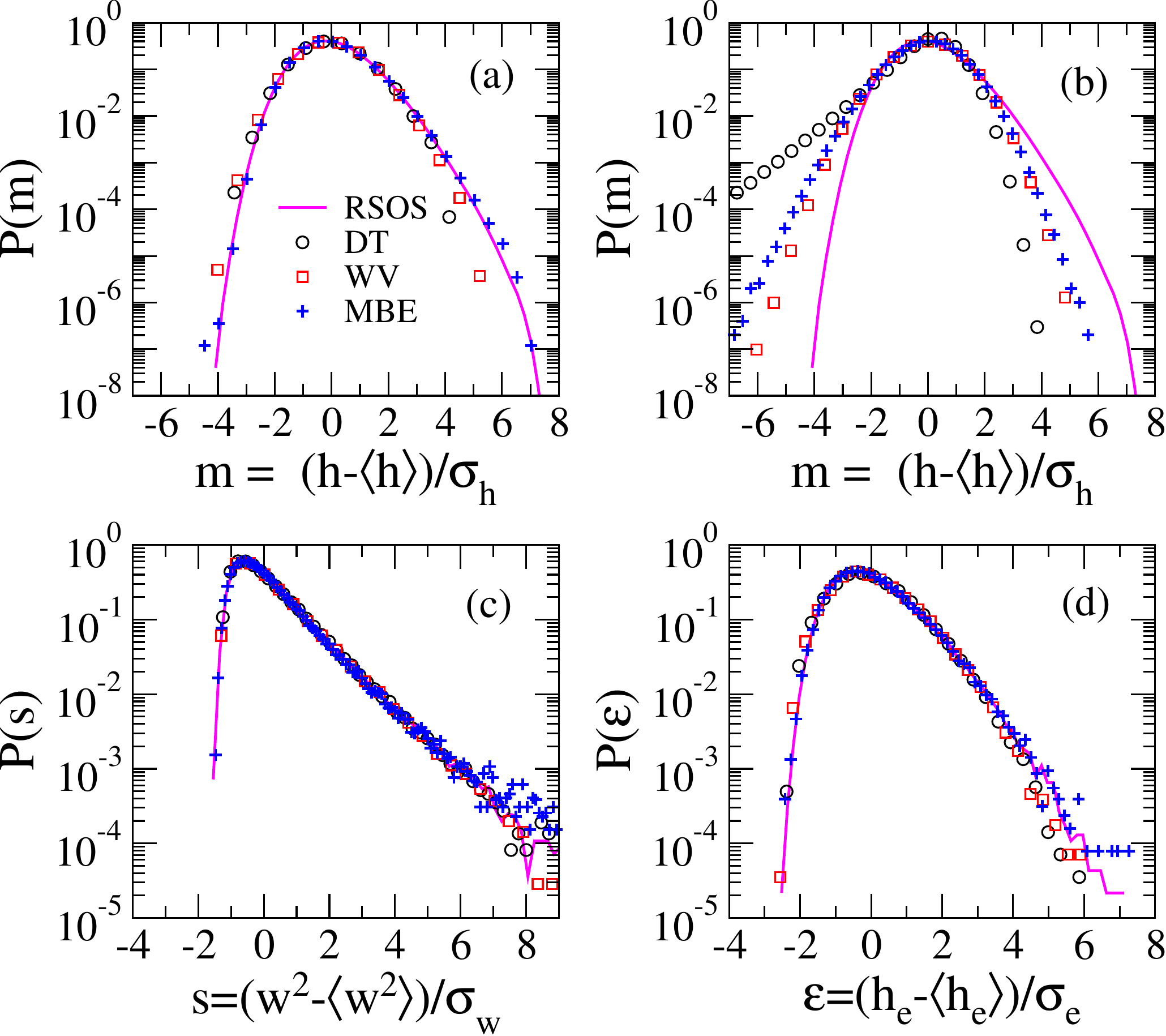}
 % dists2dModels.pdf: 754x348 pixel, 72dpi, 26.60x12.28 cm, bb=0 0 754 348
 \caption{(Color online) (a) Height,  (c) local roughness, and (d) extremal
 height probability distributions scaled to mean 0 and variance 1 for scanned
 surfaces of distinct models. The height distributions for original surfaces are
 shown in panel (b). The solid curves correspond to the KPZ surface in all
 panels. A parabolic tip of radius $R_T=16$ was used in DT and WV  and $R_T=4$
 for MBE surfaces. {Other tip sizes were checked and the results did not change.
 } For all models, 120 monolayers were deposited on $2048\times 2048$ square lattices.}
 \label{fig:dis2dmodels}.
\end{figure}

We constructed the scanned surface $\tilde{h}(\mathbf{x})$ following Lechenault
\textit{et al}.~\cite{Lechenault}:
$\tilde{h}(\mathbf{x})=\max_{\mathbf{x}'}[h(\mathbf{x}') -
g_R(\mathbf{x}-\mathbf{x}')]$, where $g_R$ gives the shape of the tip and $R_T$
its radius of curvature. We present the results for a parabolic shape
$g_R=|\mathbf{x}|^\gamma/2R_T^{\gamma-1}$, with $\gamma=2$ frequently found in
real tips~\cite{Lechenault}. We checked different tips (square that corresponds
to  $\gamma\gg 1$, semi-circle and other values of $\gamma>2$) and the central
conclusions were independent of the shape.

Experimental surfaces  analyzed with SPM are two-dimensional. So, we will
present the results for $d=2+1$ remarking that similar conclusions hold for
$d=1+1$. The effect depicted in the top insertion of Fig.~\ref{fig:phMBE2d}(c)
is enhanced in $d=2+1$ since both longitudinal and transversal directions
contribute to the growth excess. Original and scanned  surfaces for the MBE
model are compared in Figs.~\ref{fig:phMBE2d}(a) and (b). Specification of high
quality commercial AFM tips provides $R_T\lesssim 10$ nm. Assuming a typical
lattice constant of semiconductor materials as 0.5 nm, the tip sizes used in
Fig.~\ref{fig:phMBE2d} are at least as sharp as the best commercial tips. It is
worth to note that a tip radius specification refers only to the sharpest part
of the setup while its bulk, that  can  also interact with the surface depending
on the involved lengths, is much larger. The height distributions shifted to
mean zero and scaled to variance 1, $m=(h-\lrangle{h})/\sigma_h$ where
$\sigma_h=\sqrt{\lrangle{h^2}_c}$, are shown in Fig.~\ref{fig:phMBE2d}(c) for
the MBE model. This approach is widely used to evaluate the universality of
distributions in numerics and experiments
(e.g.~\cite{Racz,Palasantzas02,MF2014,almeida2013,Almeida2015,Alves11} and
references therein). The height distributions for the original surfaces deviate
from the KPZ one, which is here represented by the restricted solid-on-solid
(RSOS) model~\cite{kk,Alves14}, in both tails  and peak. However, for tips of
sizes in the range $R_T=2$ to 8, the distributions of the scanned surfaces are
remarkably close to the KPZ one.  For $R_T\gtrsim 10$ the tip is larger than the
typical mound radius ($\xi\approx 5$),  the scanned surfaces become smoother and
the right tail starts to deviate from the KPZ curve, but it is still consistent
with KPZ at least up to $10^{-4}$.  Analogous results were obtained for WV and
DT models, as illustrated in Figs.~\ref{fig:dis2dmodels}(a) and (b), with a
difference that larger tips provide the better accordance with KPZ. In case of
WV and DT surfaces, we observed the agreement with KPZ distributions in a
narrower range of tip sizes where $R_T$ is close to $\xi$. However, in a mounded
structure with high aspect ratio (height/width), the bulk of the tip can
contribute to increase the lateral width of the scanned profile. In MBE model,
for example, the original surfaces present mounds of high aspect ratio and the
KPZ distribution is still observed for a relatively tiny tip of diameter $R_T=2$
which is considerably smaller than the typical mound radius of about $\xi\approx
5$.

\begin{figure}[hb]
\centering
~~~~~MBE ~~~~~~~~~~~~~~~~~~~~~~~~~~~~~~WV\\
\includegraphics[width=8cm]{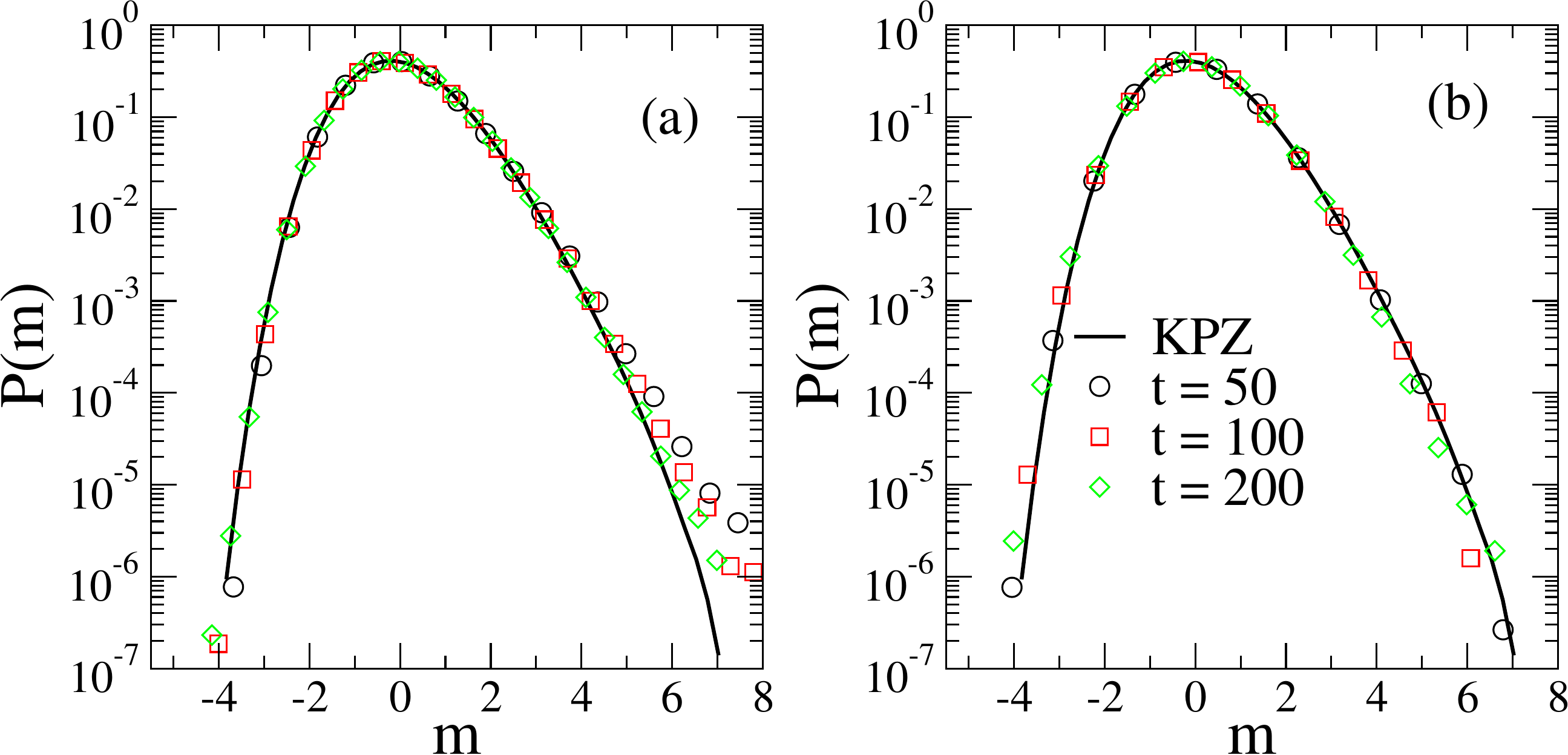}\\
\includegraphics[width=8cm]{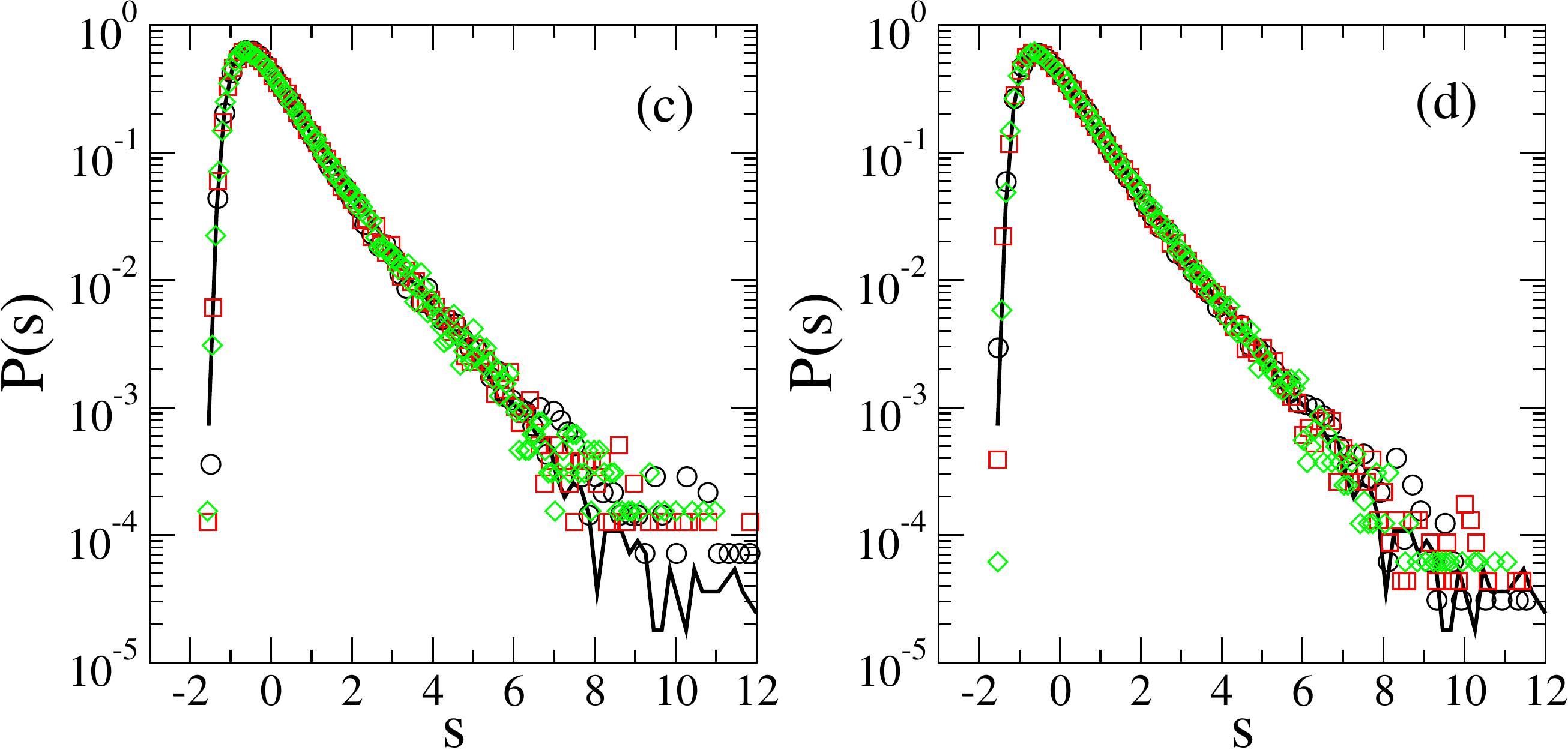}\\
\includegraphics[width=8cm]{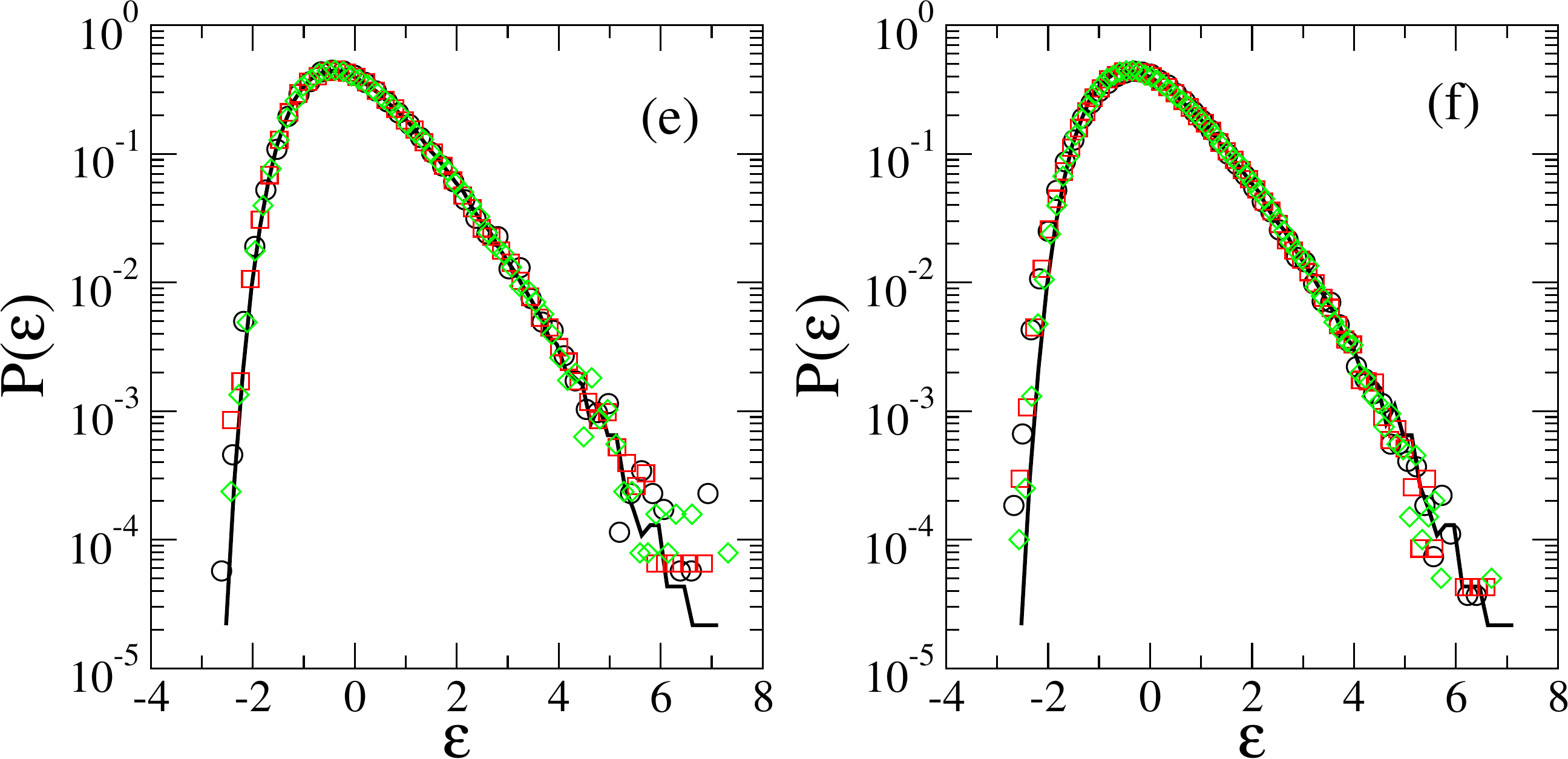}
\caption{(Color online) Scaled distributions at different times for MBE (left)
and WV (right) models in $d=2+1$ dimensions.  The height distributions were
rescaled to mean zero and variance 1. Height (top), squared roughness (middle),
and extremal height (bottom) distributions are shown. The system size is
$L=2048$ and the tip sizes are $R_T=4$ and $R_T=8$ for MBE and WV models,
respectively. The KPZ distributions (RSOS) are also shown.}
\label{fig:ph_tempos}
\end{figure}

Skewness and kurtosis are highly dependent on the tails and, therefore, are key
parameters to determine the universality of the distributions. Experimentally,
it is very  hard to obtain accurate estimates of them due to the limited
resolution in the tails of the height distributions. To the best of our
knowledge, experimental systems providing estimates of both skewness and
kurtosis  in clear agreement with KPZ are restricted to
$d=1+1$~\cite{TakeuchiPRL,TakeuchiSP,yunker} and used video microscopy to
acquire the interfaces being, therefore, rid of the mechanism investigated in
the present work. For SPM analyses, positive skewness~\cite{MF2014} or
estimates  with large uncertainties~\cite{almeida2013,Almeida2015} were
reported. For a fixed growth time,  we observed that both  skewness and kurtosis
increase with the tip size, since the lowest height contributions are being
depleted, but values consistent with KPZ ($S=0.4$ to 0.5 against 0.43 and
$K=0.3$ to 0.5 against 0.35) are observed for $R_T$ comparable to $\xi$; See
also discussion on Fig.~\ref{fig:ph_tempos} for time dependence. A
related result was obtained in Ref.~\cite{Castro14}, where the profile
statistics of equipotentials, which are akin to the output profile in SPM
techniques, generated by self-affine conducting surfaces yield a large skewness
for a growth model in the regime of the linear Edwards-Wilkinson universality
class~\cite{barabasi}, for which $S=0$.

Recently, roughness~\cite{Racz,Carrasco16} and extremal
height~\cite{Raychaudhuri,Majumdar04,Rambeau10,Carrasco16} distributions inside
boxes of size  $1\ll \ell \ll L$ with open boundary conditions have been used as
signatures of the KPZ class in
experiments~\cite{almeida2013,Almeida2015,MF2014,MF2015}. In
Figs.~\ref{fig:dis2dmodels}(c) and (d), these (scaled) distributions are
compared with the KPZ ones, where boxes of size $\ell=16$ were used.  No
significant difference in the distributions was observed in the range $\ell=8$,
16 and  32 where we measured 2.03, 1.97, and 2.08, respectively, for the
skewness of the local roughness distribution of the MBE model at time $t=120$.
This values are in agreement with previous reports~\cite{MF2014,almeida2013}.
The distributions for the original surfaces are not very different from the KPZ
one (data not shown) as pointed recently in simulations of models in the
Villain-Lai-Das Sarmas universality class~\cite{Fabio2015}, but in our
simulations the distributions for scanned surfaces are indistinguishable from
KPZ.  It is worth noticing that these distributions vary less with the tip size
than the height distribution.

The comparison of the distributions of scanned surfaces  at different
times with KPZ can be seen in Fig.~\ref{fig:ph_tempos}, showing that the results
are visually robust withing this time window. One interesting feature observed
for MBE is that the skewness initially approaches the KPZ value
$S_\mathrm{KPZ}=0.43$ as time increases in analogy with the finite-time
corrections observed in experiments~\cite{TakeuchiJSP,TakeuchiPRL,TakeuchiSP}
and simulations~\cite{Alves13,Oliveira12,Oliveira13,Alves11} of actual KPZ
systems. For example, the skewness of the height distributions in MBE are
$S=0.55,~0.49,~0.45,$ and $0.41$ for $t=50,~100,~200$ and $500$ but, differently
from actual KPZ systems, it keeps decreasing and start to deviate considerably
from KPZ for longer times. Actually, we expect that it must converge to
$S_{MBE}=0$ for very long times due to mounds increasing approximately
independently~\cite{AlvesMoreira} are expected to render a normal distribution.
For WV and DT the values move away  since shorter times and must converge to
lower values $S\approx0$ because {WV and DT models asymptotically exhibit mounded
morphology in $d=2+1$~\cite{Chatraphorn}.} Thus, the KPZ trait produced by the
finite-size tip is a transient behavior which is relevant  when the surface
characteristic lengths  are comparable with the tip size. However, for the MBE
model, the aspect ratio of the mounds initially increases with time, producing
deeper valleys and increasing the excess growth introduced by the tip
convolution and thus the KPZ hallmarks are enhanced as time evolves. Note that
the origin of the mounded morphology is the  step barrier to downhill
diffusion~\cite{Leal11a,evans2006} and its existence has been proposed in many
experiments involving thin film growth of metals,
semiconductors~\cite{evans2006} and organic materials~\cite{Zorba,Hlawacek108}.

\begin{figure}[t]
\centering
\includegraphics[width=8.5cm]{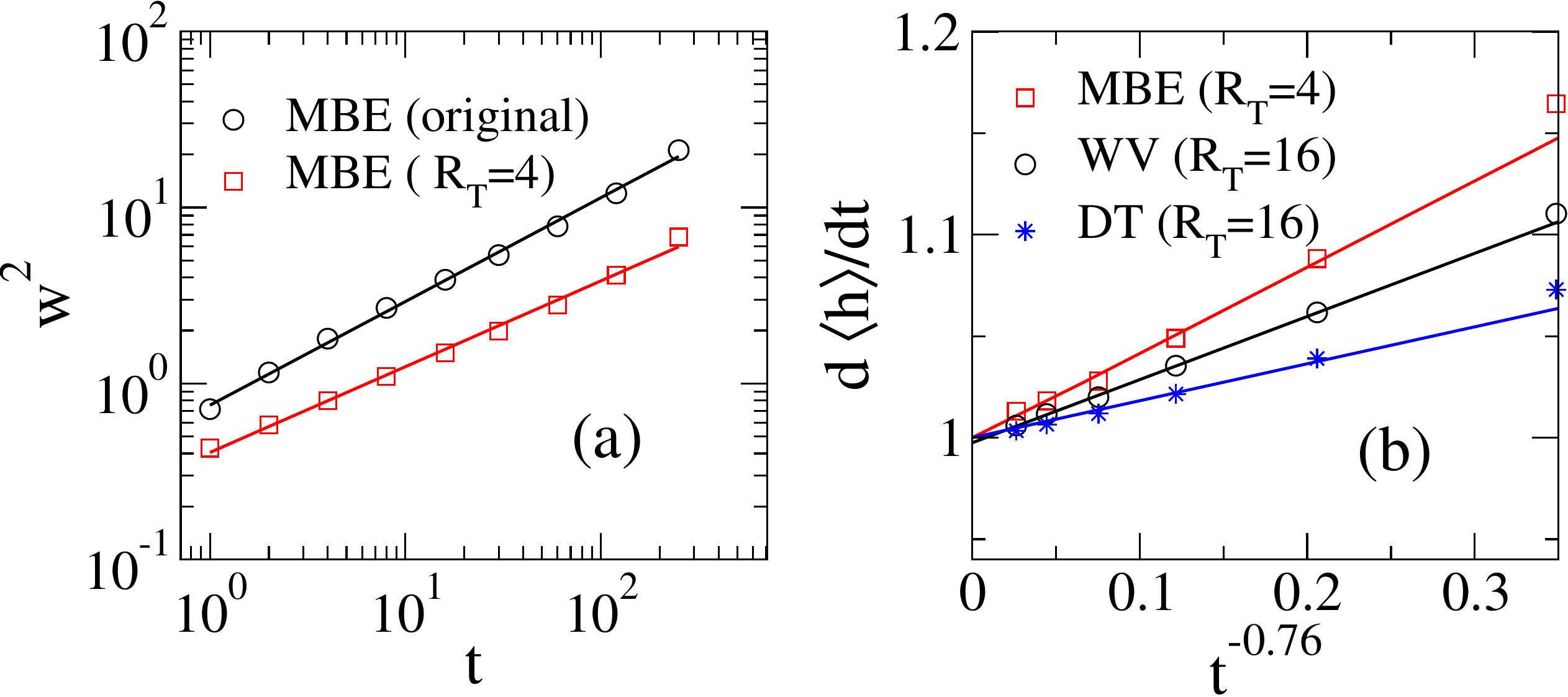}
\caption{(Color online) (a) Squared interface width against time for scanned and
original surfaces of $2+1$-MBE model. (b) Interface velocity against
$t^{-1+\beta_{kpz}}$ for three models. The solid lines are linear regressions.}
\label{fig:w2vinfMBE2d}
\end{figure}
We also analyzed the kinetic roughening during a short time interval ($\approx$
2 decades), collecting data for 10 different times and 10 independent samples
for each time, mimicking conditions reproducible in experiments. Interface width
for scanned surfaces has a quite satisfactory scaling in time,
Fig.~\ref{fig:w2vinfMBE2d}(a), with growth exponents $\beta= 0.24$ (0.30), 0.25
(0.23), and 0.25 (0.24) for MBE, WV and DT models, respectively. The numbers in
parenthesis are the exponents for the original surfaces. Notice that in DT and
WV models, the exponents for the original surfaces at short times are expected
to be given by the linear Mullins-Herring equation~\cite{barabasi} with
$\beta=0.25$ and are, inside the numerical accuracy, indistinguishable from the
KPZ exponent $\beta\approx0.24$. However, for the original surfaces of MBE model the
growth exponent is different from the KPZ one at short time while the scanned
surfaces exhibit a growth exponent consistent with KPZ. A growth exponent closer
to KPZ in scanned than in the original surfaces ar short times is a behavior
observed in other simulations we did. For example, for the Family
model~\cite{Family} in $d=1+1$, which has an exponent $\beta\approx0.25$, we
found $\beta=0.29$ in the scanned surface that approached to the one-dimensional
KPZ exponent $\beta=1/3$.

According to the KPZ ansatz,
Eq.~\eqref{eq:kpz:ansatz}, the surface growth velocity is given by
\begin{equation}
\frac{d\lrangle{h}}{dt} = v_\infty+\mbox{sgn}(\lambda)\beta \Gamma^\beta 
\lrangle{\chi} t^{\beta-1},
\end{equation}
such that plotting the interface velocity against $t^{\beta-1}$ must provide a
linear extrapolation to $v_\infty$ if the KPZ growth exponent is used. Such
behavior is confirmed in Fig.~\ref{fig:w2vinfMBE2d}(b) for all analyzed models,
{which extrapolate to $v_{\infty}=1$}. This result is not expected for
conservative dynamics, as in the models investigated here, where $v\equiv
\mathrm{const}$ independently of time.

%------------------------------------------------

\section{Electrodeposition experiments}
\label{sec:eletro}

We also investigated an experimental counterpart of this problem using
electrodeposition of iron on silicon substrates. Electrodeposition is a
electro-convective, diffusive and, therefore, nonlocal process~\cite{Sagues2000}
and is not expected to belong to the KPZ class. Indeed,  electrodeposition of
cooper~\cite{lafouresse} and iron~\cite{Cordoba-Torres} thick films revealed
anomalous (non self-affine) scaling. Profiles were acquired using AFM under both
contact and tap modes and using scanning electron microscopy (SEM), the last one
rendering surfaces rid of tip effects. {We used PPP-NCSTR
Nanosensors$^\mathrm{TM}$ tips with nominal radius of less than 10~nm  and
height of 10-15~$\mu$m.}

The iron films were grown on lightly-doped 1k$\Omega\cdot$cm, n-type (100)
silicon substrate in an area of about 0.5~cm$^2$ after native oxide cleaning by
Hidrofluoric acid. Preparation of the bath solution for the electrodeposition,
(containing 0.5~M Fe$_2$SO$_4$ and 0.5~M NaSO$_4$), followed the protocol
described in the literature~\cite{cloclo} with controlled pH = 2.5 and using a
potentiostatic mode under a potential of $-1.8$~V. AFM and SEM images of
$4\times 4~\mu$m$^2$ with a resolution of $1024\times 1024$ pixels were
acquired. The SEM images were processed with a standard blur (low-pass) filter
to erase the effects of electric accumulation in the grain  borders. The
gray-scale of SEM images was linearly converted to  a height interval $0\le
\tilde{h}\le 1$~\cite{Castaneda}. Even though the vertical resolution cannot be
obtained with SEM, the gray-scale method has been used to estimate fractal
dimensions in semiconductor surfaces~\cite{Castaneda}. For our purpose, the
actual height scale does not matter since distribution is shifted to zero mean
and variance 1. Possibly the linear assumption in gray-scale method is a rough
approximation but it is enough to show that the surface distribution is
inconsistent with KPZ.

\begin{figure}[h]
 \centering
 ~~~~~~~~\includegraphics[width=3.25cm]{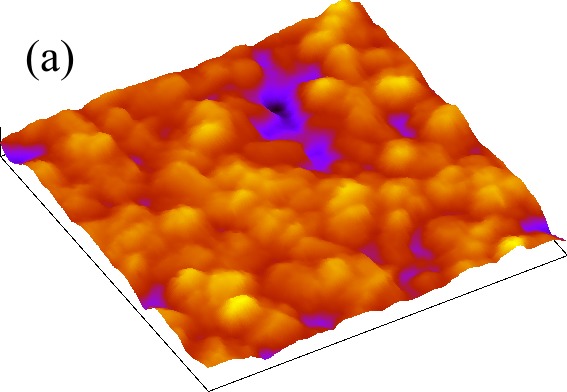} ~~~~~ \includegraphics[width=3cm]{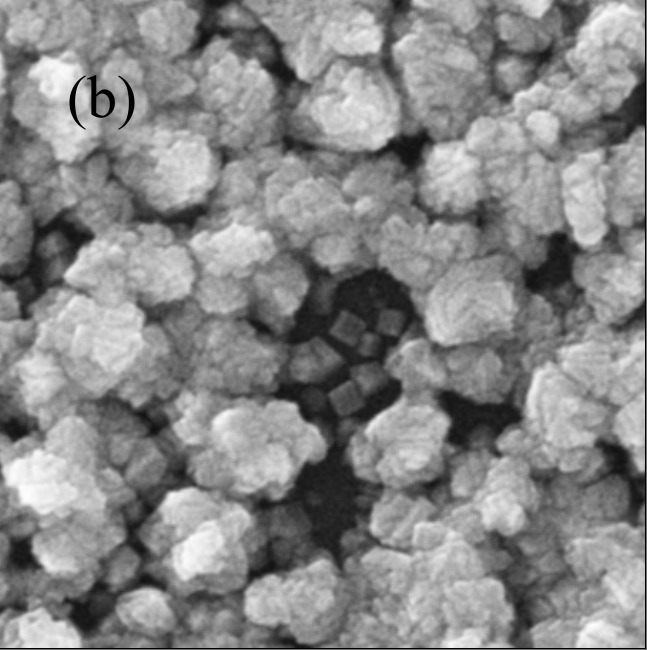}\\
  \includegraphics[width=8cm]{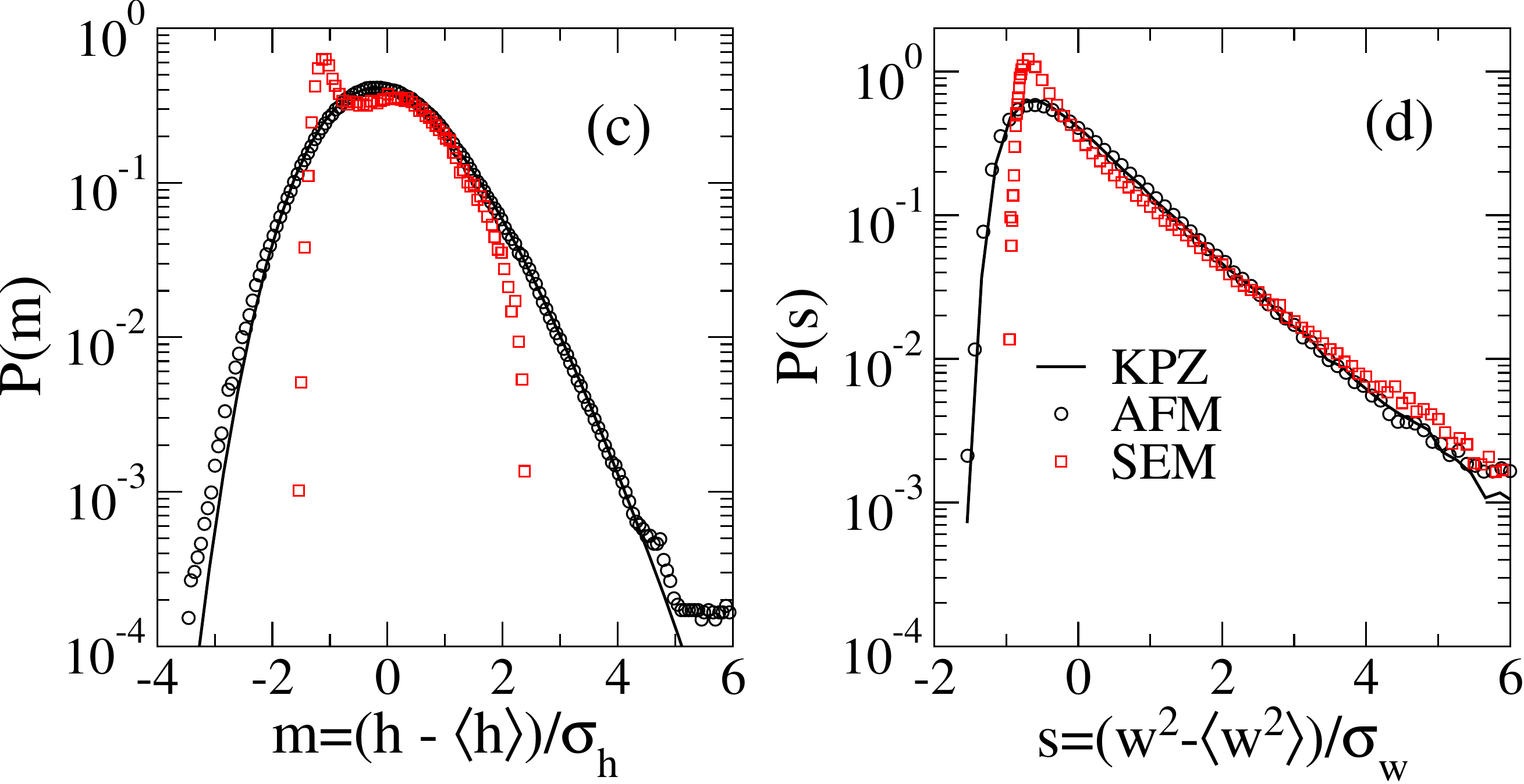}\\
% ph_t150.pdf: 754x388 pixel, 72dpi, 26.60x13.69 cm, bb=0 0 754 388
\caption{(Color online)\label{fig:exp}
{4$\times$4 $\mu$m$^2$} images of iron films after a growth time $t=150$ s
obtained with (a) AFM  in contact mode and (b) SEM imaging techniques. (c)
Height  and (d) local roughness distributions scaled to mean zero and variance 1
obtained with AFM and SEM are compared with the KPZ class.}
\end{figure}

Figures~\ref{fig:exp}(a) and (b) show a film surface for $t=150$s generated by
AFM (contact mode) and SEM, respectively. Five samples  were used
for statistics. The SEM image reveals a surface with large grains ($\sim$300 nm)
where the substrate still can be resolved, in contrast with smoothed surface
produced by the AFM technique. The film approximate thickness  is $\sim500$ nm
implying in a high aspect ratio. The rescaled height distributions are compared
with the KPZ one in Fig.~\ref{fig:exp}(c), in which one can see that the SEM
provides a bimodal distribution whereas the smoothed AFM surfaces are very close
to the KPZ distribution. The bimodal distribution reflects the co-existence of
small and large grains, the former having lower growth rates due to the
screening effects caused by the latter. Local roughness [Fig.~\ref{fig:exp}(d)]
and extremal height (data not shown) distributions of the AFM images are not
distinguishable from the KPZ curves whereas those for SEM are. Here we have an
example where the tip effect is relevant even when the typical lateral length is
much larger than  the tip specification. No significant difference between
contact and tap modes was observed.

%------------------------------------------------

\section{Ruling  out false positives}
\label{sec:discuss}

Summed up the observations reported in the previous section, it is worth to point out
strategies to rid the analysis from this tip induced KPZ traits.
There exist other KPZ hallmarks that could be used to confirm the
universality as, for example, temporal and spatial correlation functions
(see, e.g., \cite{TakeuchiJSP,MF2014,odor2014aging,Carrasco} and
references therein). The former requires a huge amount of data for a
reliable analysis and therefore are not currently feasible in experiments
with SPM. For the latter, a good agreement between  experimental data for
organic films and $2+1$-KPZ models was reported~\cite{MF2014}. The
spatial covariance is given by
\begin{equation}
C_s(r)=\lrangle{h(\mathbf{x}+\mathbf{r})h(\mathbf{x})}-\lrangle{h^2}\simeq(\Gamma t)^{2\beta}
G\left[\frac{A_hr^{2\alpha}}{2(\Gamma t)^{2\beta}}\right]
\end{equation}
where the brackets represent average over positions and samples, $G(x)$ is a scaling
function and $A_h$ a phenomenological parameter to be determined. The parameter
$A_h$ of the spatial correlators could, in principle, be determined using the
Krug-Meakin method~\cite{Krug90}. However, we follow the recipe of
Ref.~\cite{MF2014}, suited to experimental data: we plot $G(u)=C_s/(\Gamma
t)^{2\beta}$ against $u=\frac{1}{2}A_h r^{2\alpha}/(\Gamma t)^{2\beta}$ and fit
$\Gamma$ and $A_h$ to provide $G(0)=\lrangle{\chi^2}_c=0.237$ with a slope -1
for $u\approx 0$. For  $r \ll R_T$ the surface is essentially smooth and,
therefore, we excluded $r<R_T/2$ from the fit procedure. The parameters were
therefore extracted fitting the data in range $R_T/2<r<R_T$ to extrapolate to
$\lrangle{\chi^2}_c$ with a slope -1 at $u=0$. If the rescaling is performed
with the accepted KPZ exponent $\alpha=0.39$~\cite{Kelling}, the estimates for
scanned  surfaces are $A_h\approx 0.94$ and $(\Gamma t)^{2\beta} \approx 11.9$
for MBE  and $A_h\approx0.32$ and $(\Gamma t)^{2\beta} \approx 5$ for WV models.
For the electrodeposition experiments, the values are $A_h\approx 90$ and 
$(\Gamma t)^{2\beta}\approx 1210$.
\begin{figure}[t]
 \centering
 \includegraphics[width=8cm]{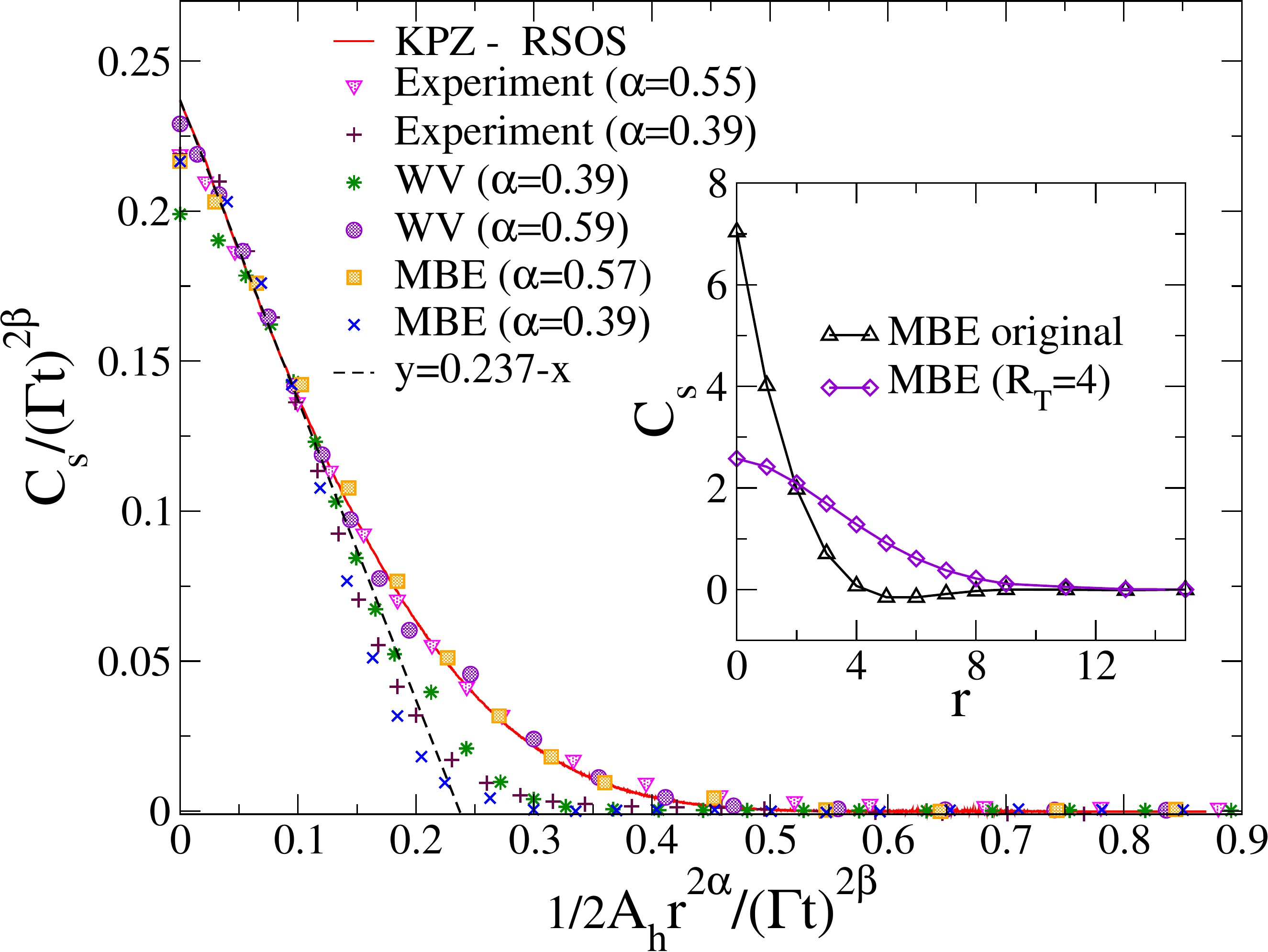}
 % covwv.pdf: 736x520 pixel, 72dpi, 25.96x18.34 cm, bb=0 0 736 520
\caption{(Color online) Scaled spatial covariance for MBE and WV surfaces
scanned with probe tips of sizes $R_T=4$ and $R_T=16$, respectively, are shown.
Two values of $\alpha$ were used: theoretically conjectured for KPZ and directly
measured by the best data collapse. The covariance for the KPZ class is also
shown. Inset shows the original and scanned surfaces for MBE model without
rescaling showing the mounded surface signature (minimum in $C_s$) for the
former but not for the latter. {The error bars are slightly larger
(experiments) or smaller (simulations) than symbols and were omitted for sake of
clearness.}} \label{fig:covwv}
\end{figure}

Figure~\ref{fig:covwv} compares the scaled spatial covariance of MBE and WV
models and electrodeposition films with the $2+1$-KPZ curve. The obtained
correlation functions clearly deviate from the curve expected for the KPZ class
if the accepted KPZ exponent~\cite{Kelling} $\alpha=0.39$ is used, indicating
that this analysis succeeded the test and does not provide a false positive.
We could also simultaneously fit $\alpha$ looking for the best
collapse. In this case, exponents larger than the KPZ one ($\alpha=0.57$,
$\alpha=0.59$, 0.55 for MBE, WV, and experiments respectively) are obtained
pointing out again that the system is not KPZ. Notice, however, that a very nice
collapse with the KPZ curve is found when the non-KPZ $\alpha$ exponents are
used, Fig.~\ref{fig:covwv}, configuring a potential false positive.

Even though the growth exponent $\beta$ approaches the KPZ class after
scanning and could potentially induce a false positive, in many situations the
value can still be far from KPZ, as aforementioned for Family
model~\cite{Family} in $d=1+1$. So, the roughness and growth exponents, in the
canonical way~\cite{barabasi}, are the basic starting points for the
verification of the KPZ class. Other measurements such as distributions and
correlators are fundamental to complement the analysis.

Other possible path is to analyze the actual distribution of $\chi$ without the
scaling to mean 0 and variance 1, \textit{i e}., as performed in the
experimental realization of $1+1$-KPZ class~\cite{TakeuchiPRL,TakeuchiSP}. For
example, curved and flat KPZ subclasses are, in practice, indistinguishable if
this null mean and unity variance strategy is used;  See Fig. 6 of
Ref.~\cite{takeuchi2014}, for example. Alternatively, one could fix either
average or variance to unity (not to zero) extracting all remaining  KPZ
properties~\cite{Alves13,Oliveira13}. The slope of plots $d\lrangle{h}/dt$
versus $t^{\beta-1}$ yields
$g_1=\mbox{sgn}(\lambda)\beta\Gamma^\beta\lrangle{\chi}$ whereas
$g_2=\lrangle{h^2}_c/t^{2\beta}$ yields $\Gamma^{2\beta}\lrangle{\chi^2}_c$ such
that the ratio $R=\beta^2 g_2/g_1=\lrangle{\chi^2}_c/\lrangle{\chi}^2\approx
0.32$  is a universal KPZ value~\cite{Oliveira13}. We obtained $R\approx 0.13$,
0.09 and 0.15 for MBE, WV and DT models, which are not consistent with the
$2+1$-KPZ class. Note that the scanned surfaces have a positively skewed height
distributions that implies $\lambda>0$. However, at the light of the KPZ ansatz,
the positive slope Fig.~\ref{fig:w2vinfMBE2d}(b) implies that $\lrangle{\chi}>0$
contrasting with negative $2+1$-KPZ value of $-0.83$~\cite{MF12,Alves14}.

\section{Conclusions}
\label{sec:conclusions}

In summary, we provide evidences that SPM can artificially introduce traits of
the KPZ class in systems where they are not expected. The false KPZ hallmarks
were observed in  models when the tip convolution introduces a non-negligible
excess growth in the original surface. This situation was more clearly observed
when the tip specification radius was comparable with the characteristic
correlation lengths of rough surfaces or when the surfaces have mounds with a
high aspect ratio (height/width) due to the interaction with the tip bulk.
Experimental results for electrodeposition of iron also support that KPZ traits
are enhance by high aspect ratios. The underlying mechanism of the KPZ features
is the excess velocity introduced by the probe tip scanning. While, on the one
hand, it was recently shown that  the surface smoothing facilitate the unveiling
of the universality in genuine KPZ systems with strong correction to the
scaling~\cite{AlvesOrigins,Alves2016}, on the other hand, it can introduce
artificial traits in dynamics different from KPZ. We also discussed ways to rule
out false positives for KPZ. It is important to remark that we observed
artificial KPZ traits  only for relatively short times and  as
$t\rightarrow\infty$ the characteristic lengths become much larger than the tip
size and the excess growth introduced by the scanning must become negligible.
However, it is possible to figure out situations where the artificial KPZ
properties can last for long times.

A possible explanation for experimental observations of the KPZ class using SPM
was given. Experimental realization of the KPZ class with SPM must  assure that
the tip probe effects are negligible.  Concerning some of the  experimental
realizations of the KPZ class in 2+1 dimensions
\cite{Palasantzas02,MF13,almeida2013,Almeida2015}, the AFM images shows that
typical characteristic lengths extracted from correlation functions are
considerably above a typical tip radius of 10-20~nm. However, a high aspect
ratio cannot be discarded using only AFM images since the deeply grooved regions
cannot be resolved with this technique. A systematic analysis of tip size or
using non-contact scanning methods such as SEM in these experimental systems
would be interesting future works.

\section*{Acknowledgements}

This work was partially supported by the Brazilian agencies CNPq and FAPEMIG. We
thank Maxmiliano Munford, Renan Almeida, and Thiago de Assis for motivating
discussions.  We also thank Rodolfo Cuerno, Tiago Oliveira and Sukarno Ferreira
by their critical opinions on the manuscript.

%\bibliography{tip}

\begin{thebibliography}{65}%
\makeatletter
\providecommand \@ifxundefined [1]{%
 \@ifx{#1\undefined}
}%
\providecommand \@ifnum [1]{%
 \ifnum #1\expandafter \@firstoftwo
 \else \expandafter \@secondoftwo
 \fi
}%
\providecommand \@ifx [1]{%
 \ifx #1\expandafter \@firstoftwo
 \else \expandafter \@secondoftwo
 \fi
}%
\providecommand \natexlab [1]{#1}%
\providecommand \enquote  [1]{``#1''}%
\providecommand \bibnamefont  [1]{#1}%
\providecommand \bibfnamefont [1]{#1}%
\providecommand \citenamefont [1]{#1}%
\providecommand \href@noop [0]{\@secondoftwo}%
\providecommand \href [0]{\begingroup \@sanitize@url \@href}%
\providecommand \@href[1]{\@@startlink{#1}\@@href}%
\providecommand \@@href[1]{\endgroup#1\@@endlink}%
\providecommand \@sanitize@url [0]{\catcode `\\12\catcode `\$12\catcode
  `\&12\catcode `\#12\catcode `\^12\catcode `\_12\catcode `\%12\relax}%
\providecommand \@@startlink[1]{}%
\providecommand \@@endlink[0]{}%
\providecommand \url  [0]{\begingroup\@sanitize@url \@url }%
\providecommand \@url [1]{\endgroup\@href {#1}{\urlprefix }}%
\providecommand \urlprefix  [0]{URL }%
\providecommand \Eprint [0]{\href }%
\providecommand \doibase [0]{http://dx.doi.org/}%
\providecommand \selectlanguage [0]{\@gobble}%
\providecommand \bibinfo  [0]{\@secondoftwo}%
\providecommand \bibfield  [0]{\@secondoftwo}%
\providecommand \translation [1]{[#1]}%
\providecommand \BibitemOpen [0]{}%
\providecommand \bibitemStop [0]{}%
\providecommand \bibitemNoStop [0]{.\EOS\space}%
\providecommand \EOS [0]{\spacefactor3000\relax}%
\providecommand \BibitemShut  [1]{\csname bibitem#1\endcsname}%
\let\auto@bib@innerbib\@empty
%</preamble>
\bibitem [{\citenamefont {Privman}(1990)}]{privman1990finite}%
  \BibitemOpen
  \bibinfo {editor} {\bibfnamefont {V.}~\bibnamefont {Privman}},\ ed.,\ \href
  {http://books.google.com.br/books?isbn=9810201087} {\emph {\bibinfo {title}
  {Finite Size Scaling and Numerical Simulation of Statistical Systems}}}\
  (\bibinfo  {publisher} {World Scientific},\ \bibinfo {address} {Singapore},\
  \bibinfo {year} {1990})\BibitemShut {NoStop}%
\bibitem [{\citenamefont {Binder}(1981)}]{binder1981finite}%
  \BibitemOpen
  \bibfield  {author} {\bibinfo {author} {\bibfnamefont {K.}~\bibnamefont
  {Binder}},\ }\bibfield  {title} {\enquote {\bibinfo {title} {Finite size
  scaling analysis of {Ising} model block distribution functions},}\ }\href
  {http://dx.doi.org/10.1007/BF01293604} {\bibfield  {journal} {\bibinfo
  {journal} {Z. Physik B - Condensed Matter}\ }\textbf {\bibinfo {volume}
  {43}},\ \bibinfo {pages} {119} (\bibinfo {year} {1981})}\BibitemShut
  {NoStop}%
\bibitem [{\citenamefont {Henkel}\ \emph {et~al.}(2008)\citenamefont {Henkel},
  \citenamefont {Hinrichsen}, \citenamefont {L{\"u}beck},\ and\ \citenamefont
  {Pleimling}}]{henkel2008non}%
  \BibitemOpen
  \bibfield  {author} {\bibinfo {author} {\bibfnamefont {M.}~\bibnamefont
  {Henkel}}, \bibinfo {author} {\bibfnamefont {H.}~\bibnamefont {Hinrichsen}},
  \bibinfo {author} {\bibfnamefont {S.}~\bibnamefont {L{\"u}beck}}, \ and\
  \bibinfo {author} {\bibfnamefont {M.}~\bibnamefont {Pleimling}},\ }\href
  {http://www.springer.com/us/book/9781402087646} {\emph {\bibinfo {title}
  {Non-equilibrium phase transitions}}},\ Vol.~\bibinfo {volume} {1}\ (\bibinfo
   {publisher} {Springer},\ \bibinfo {address} {Berlin},\ \bibinfo {year}
  {2008})\BibitemShut {NoStop}%
\bibitem [{\citenamefont {Krug}\ \emph {et~al.}(1992)\citenamefont {Krug},
  \citenamefont {Meakin},\ and\ \citenamefont {Halpin-Healy}}]{KrugPRA92}%
  \BibitemOpen
  \bibfield  {author} {\bibinfo {author} {\bibfnamefont {J.}~\bibnamefont
  {Krug}}, \bibinfo {author} {\bibfnamefont {P.}~\bibnamefont {Meakin}}, \ and\
  \bibinfo {author} {\bibfnamefont {T.}~\bibnamefont {Halpin-Healy}},\
  }\bibfield  {title} {\enquote {\bibinfo {title} {Amplitude universality for
  driven interfaces and directed polymers in random media},}\ }\href@noop {}
  {\bibfield  {journal} {\bibinfo  {journal} {Phys. Rev. A}\ }\textbf {\bibinfo
  {volume} {45}},\ \bibinfo {pages} {638} (\bibinfo {year} {1992})}\BibitemShut
  {NoStop}%
\bibitem [{\citenamefont {Kardar}\ \emph {et~al.}(1986)\citenamefont {Kardar},
  \citenamefont {Parisi},\ and\ \citenamefont {Zhang}}]{KPZ}%
  \BibitemOpen
  \bibfield  {author} {\bibinfo {author} {\bibfnamefont {M.}~\bibnamefont
  {Kardar}}, \bibinfo {author} {\bibfnamefont {G.}~\bibnamefont {Parisi}}, \
  and\ \bibinfo {author} {\bibfnamefont {Y.-C.}\ \bibnamefont {Zhang}},\
  }\bibfield  {title} {\enquote {\bibinfo {title} {Dynamic scaling of growing
  interfaces},}\ }\href {http://dx.doi.org/10.1103/PhysRevLett.56.889}
  {\bibfield  {journal} {\bibinfo  {journal} {Phys. Rev. Lett.}\ }\textbf
  {\bibinfo {volume} {56}},\ \bibinfo {pages} {889} (\bibinfo {year}
  {1986})}\BibitemShut {NoStop}%
\bibitem [{\citenamefont {Family}\ and\ \citenamefont
  {Vicsek}(1985)}]{Family85}%
  \BibitemOpen
  \bibfield  {author} {\bibinfo {author} {\bibfnamefont {F.}~\bibnamefont
  {Family}}\ and\ \bibinfo {author} {\bibfnamefont {T.}~\bibnamefont
  {Vicsek}},\ }\bibfield  {title} {\enquote {\bibinfo {title} {Scaling of the
  active zone in the {Eden} process on percolation networks and the ballistic
  deposition model},}\ }\href {http://dx.doi.org/10.1088/0305-4470/18/2/005}
  {\bibfield  {journal} {\bibinfo  {journal} {J. Phys. A: Math. Gen.}\ }\textbf
  {\bibinfo {volume} {18}},\ \bibinfo {pages} {L75} (\bibinfo {year}
  {1985})}\BibitemShut {NoStop}%
\bibitem [{\citenamefont {Barabasi}\ and\ \citenamefont
  {Stanley}(1995)}]{barabasi}%
  \BibitemOpen
  \bibfield  {author} {\bibinfo {author} {\bibfnamefont {A.-L.}\ \bibnamefont
  {Barabasi}}\ and\ \bibinfo {author} {\bibfnamefont {H.~E.}\ \bibnamefont
  {Stanley}},\ }\href {https://books.google.com.br/books?id=W4SqcNr8PLYC}
  {\emph {\bibinfo {title} {Fractal Concepts in Surface Growth}}}\ (\bibinfo
  {publisher} {Cambridge University Press},\ \bibinfo {address} {Cambridge,
  England},\ \bibinfo {year} {1995})\BibitemShut {NoStop}%
\bibitem [{\citenamefont {Pr\"ahofer}\ and\ \citenamefont
  {Spohn}(2000)}]{PraSpo1}%
  \BibitemOpen
  \bibfield  {author} {\bibinfo {author} {\bibfnamefont {M.}~\bibnamefont
  {Pr\"ahofer}}\ and\ \bibinfo {author} {\bibfnamefont {H.}~\bibnamefont
  {Spohn}},\ }\bibfield  {title} {\enquote {\bibinfo {title} {Universal
  distributions for growth processes in $1+1$ dimensions and random
  matrices},}\ }\href {http://dx.doi.org/10.1103/PhysRevLett.84.4882}
  {\bibfield  {journal} {\bibinfo  {journal} {Phys. Rev. Lett.}\ }\textbf
  {\bibinfo {volume} {84}},\ \bibinfo {pages} {4882} (\bibinfo {year}
  {2000})}\BibitemShut {NoStop}%
\bibitem [{\citenamefont {Johansson}(2000)}]{Johansson-CMP2000}%
  \BibitemOpen
  \bibfield  {author} {\bibinfo {author} {\bibfnamefont {K.}~\bibnamefont
  {Johansson}},\ }\bibfield  {title} {\enquote {\bibinfo {title} {Shape
  fluctuations and random matrices},}\ }\href
  {http://link.springer.com/article/10.1007/s002200050027} {\bibfield
  {journal} {\bibinfo  {journal} {Commun. Math. Phys.}\ }\textbf {\bibinfo
  {volume} {209}},\ \bibinfo {pages} {437} (\bibinfo {year}
  {2000})}\BibitemShut {NoStop}%
\bibitem [{\citenamefont {Sasamoto}\ and\ \citenamefont
  {Spohn}(2010)}]{SasaSpo1}%
  \BibitemOpen
  \bibfield  {author} {\bibinfo {author} {\bibfnamefont {T.}~\bibnamefont
  {Sasamoto}}\ and\ \bibinfo {author} {\bibfnamefont {H.}~\bibnamefont
  {Spohn}},\ }\bibfield  {title} {\enquote {\bibinfo {title} {One-dimensional
  {Kardar-Parisi-Zhang} equation: {An} exact solution and its universality},}\
  }\href {http://dx.doi.org/10.1103/PhysRevLett.104.230602} {\bibfield
  {journal} {\bibinfo  {journal} {Phys. Rev. Lett.}\ }\textbf {\bibinfo
  {volume} {104}},\ \bibinfo {pages} {230602} (\bibinfo {year}
  {2010})}\BibitemShut {NoStop}%
\bibitem [{\citenamefont {Takeuchi}\ and\ \citenamefont
  {Sano}(2010)}]{TakeuchiPRL}%
  \BibitemOpen
  \bibfield  {author} {\bibinfo {author} {\bibfnamefont {K.~A.}\ \bibnamefont
  {Takeuchi}}\ and\ \bibinfo {author} {\bibfnamefont {M.}~\bibnamefont
  {Sano}},\ }\bibfield  {title} {\enquote {\bibinfo {title} {Universal
  fluctuations of growing interfaces: {Evidence} in turbulent liquid
  crystals},}\ }\href {http://dx.doi.org/10.1103/PhysRevLett.104.230601}
  {\bibfield  {journal} {\bibinfo  {journal} {Phys. Rev. Lett.}\ }\textbf
  {\bibinfo {volume} {104}},\ \bibinfo {pages} {230601} (\bibinfo {year}
  {2010})}\BibitemShut {NoStop}%
\bibitem [{\citenamefont {Calabrese}\ and\ \citenamefont
  {Le~Doussal}(2011)}]{Calabrese}%
  \BibitemOpen
  \bibfield  {author} {\bibinfo {author} {\bibfnamefont {P.}~\bibnamefont
  {Calabrese}}\ and\ \bibinfo {author} {\bibfnamefont {P.}~\bibnamefont
  {Le~Doussal}},\ }\bibfield  {title} {\enquote {\bibinfo {title} {Exact
  solution for the {Kardar-Parisi-Zhang} equation with flat initial
  conditions},}\ }\href {http://dx.doi.org/10.1103/PhysRevLett.106.250603}
  {\bibfield  {journal} {\bibinfo  {journal} {Phys. Rev. Lett.}\ }\textbf
  {\bibinfo {volume} {106}},\ \bibinfo {pages} {250603} (\bibinfo {year}
  {2011})}\BibitemShut {NoStop}%
\bibitem [{\citenamefont {Oliveira}\ \emph {et~al.}(2012)\citenamefont
  {Oliveira}, \citenamefont {Ferreira},\ and\ \citenamefont
  {Alves}}]{Oliveira12}%
  \BibitemOpen
  \bibfield  {author} {\bibinfo {author} {\bibfnamefont {T.~J.}\ \bibnamefont
  {Oliveira}}, \bibinfo {author} {\bibfnamefont {S.~C.}\ \bibnamefont
  {Ferreira}}, \ and\ \bibinfo {author} {\bibfnamefont {S.~G.}\ \bibnamefont
  {Alves}},\ }\bibfield  {title} {\enquote {\bibinfo {title} {Universal
  fluctuations in {Kardar-Parisi-Zhang} growth on one-dimensional flat
  substrates},}\ }\href {http://dx.doi.org/10.1103/PhysRevE.85.010601}
  {\bibfield  {journal} {\bibinfo  {journal} {Phys. Rev. E}\ }\textbf {\bibinfo
  {volume} {85}},\ \bibinfo {pages} {010601} (\bibinfo {year}
  {2012})}\BibitemShut {NoStop}%
\bibitem [{\citenamefont {Takeuchi}\ \emph {et~al.}(2011)\citenamefont
  {Takeuchi}, \citenamefont {Sano}, \citenamefont {Sasamoto},\ and\
  \citenamefont {Spohn}}]{TakeuchiSP}%
  \BibitemOpen
  \bibfield  {author} {\bibinfo {author} {\bibfnamefont {K.~A.}\ \bibnamefont
  {Takeuchi}}, \bibinfo {author} {\bibfnamefont {M.}~\bibnamefont {Sano}},
  \bibinfo {author} {\bibfnamefont {T.}~\bibnamefont {Sasamoto}}, \ and\
  \bibinfo {author} {\bibfnamefont {H.}~\bibnamefont {Spohn}},\ }\bibfield
  {title} {\enquote {\bibinfo {title} {Growing interfaces uncover universal
  fluctuations behind scale invariance},}\ }\href
  {http://www.nature.com/articles/srep00034} {\bibfield  {journal} {\bibinfo
  {journal} {Sci. Rep.}\ }\textbf {\bibinfo {volume} {1}},\ \bibinfo {pages}
  {34} (\bibinfo {year} {2011})}\BibitemShut {NoStop}%
\bibitem [{\citenamefont {Tracy}\ and\ \citenamefont {Widom}(1994)}]{TW}%
  \BibitemOpen
  \bibfield  {author} {\bibinfo {author} {\bibfnamefont {C.}~\bibnamefont
  {Tracy}}\ and\ \bibinfo {author} {\bibfnamefont {H.}~\bibnamefont {Widom}},\
  }\bibfield  {title} {\enquote {\bibinfo {title} {Level-spacing distributions
  and the {Airy} kernel},}\ }\href {http://dx.doi.org/10.1007/BF02100489}
  {\bibfield  {journal} {\bibinfo  {journal} {Commun. Math. Phys.}\ }\textbf
  {\bibinfo {volume} {159}},\ \bibinfo {pages} {151} (\bibinfo {year}
  {1994})}\BibitemShut {NoStop}%
\bibitem [{\citenamefont {Halpin-Healy}(2012)}]{MF12}%
  \BibitemOpen
  \bibfield  {author} {\bibinfo {author} {\bibfnamefont {T.}~\bibnamefont
  {Halpin-Healy}},\ }\bibfield  {title} {\enquote {\bibinfo {title}
  {($2\mathbf{+}1$)-dimensional directed polymer in a random medium: {S}caling
  phenomena and universal distributions},}\ }\href@noop {} {\bibfield
  {journal} {\bibinfo  {journal} {Phys. Rev. Lett.}\ }\textbf {\bibinfo
  {volume} {109}},\ \bibinfo {pages} {170602} (\bibinfo {year}
  {2012})}\BibitemShut {NoStop}%
\bibitem [{\citenamefont {Oliveira}\ \emph {et~al.}(2013)\citenamefont
  {Oliveira}, \citenamefont {Alves},\ and\ \citenamefont
  {Ferreira}}]{Oliveira13}%
  \BibitemOpen
  \bibfield  {author} {\bibinfo {author} {\bibfnamefont {T.~J.}\ \bibnamefont
  {Oliveira}}, \bibinfo {author} {\bibfnamefont {S.~G.}\ \bibnamefont {Alves}},
  \ and\ \bibinfo {author} {\bibfnamefont {S.~C.}\ \bibnamefont {Ferreira}},\
  }\bibfield  {title} {\enquote {\bibinfo {title} {{Kardar-Parisi-Zhang}
  universality class in ($2+1$) dimensions: Universal geometry-dependent
  distributions and finite-time corrections},}\ }\href
  {http://dx.doi.org/10.1103/PhysRevE.87.040102} {\bibfield  {journal}
  {\bibinfo  {journal} {Phys. Rev. E}\ }\textbf {\bibinfo {volume} {87}},\
  \bibinfo {pages} {040102} (\bibinfo {year} {2013})}\BibitemShut {NoStop}%
\bibitem [{\citenamefont {Halpin-Healy}(2013)}]{MF13}%
  \BibitemOpen
  \bibfield  {author} {\bibinfo {author} {\bibfnamefont {T.}~\bibnamefont
  {Halpin-Healy}},\ }\bibfield  {title} {\enquote {\bibinfo {title} {Extremal
  paths, the stochastic heat equation, and the three-dimensional
  {Kardar-Parisi-Zhang} universality class},}\ }\href@noop {} {\bibfield
  {journal} {\bibinfo  {journal} {Phys. Rev. E}\ }\textbf {\bibinfo {volume}
  {88}},\ \bibinfo {pages} {042118} (\bibinfo {year} {2013})}\BibitemShut
  {NoStop}%
\bibitem [{\citenamefont {Alves}\ and\ \citenamefont
  {Ferreira}(2012)}]{Alves12}%
  \BibitemOpen
  \bibfield  {author} {\bibinfo {author} {\bibfnamefont {S.~G.}\ \bibnamefont
  {Alves}}\ and\ \bibinfo {author} {\bibfnamefont {S.~C.}\ \bibnamefont
  {Ferreira}},\ }\bibfield  {title} {\enquote {\bibinfo {title} {{E}den
  clusters in three dimensions and the {Kardar-Parisi-Zhang} universality
  class},}\ }\href {http://dx.doi.org/10.1088/1742-5468/2012/10/P10011}
  {\bibfield  {journal} {\bibinfo  {journal} {J. Stat. Mech.}\ }\textbf
  {\bibinfo {volume} {2012}},\ \bibinfo {pages} {P10011} (\bibinfo {year}
  {2012})}\BibitemShut {NoStop}%
\bibitem [{\citenamefont {Carrasco}\ and\ \citenamefont
  {Oliveira}(2016{\natexlab{a}})}]{Carrasco2016b}%
  \BibitemOpen
  \bibfield  {author} {\bibinfo {author} {\bibfnamefont {I.~S.~S.}\
  \bibnamefont {Carrasco}}\ and\ \bibinfo {author} {\bibfnamefont {T.~J.}\
  \bibnamefont {Oliveira}},\ }\bibfield  {title} {\enquote {\bibinfo {title}
  {{Universality and geometry dependence in the class of the nonlinear
  molecular beam epitaxy equation}},}\ }\href {http://arxiv.org/abs/1606.06097}
  {\  (\bibinfo {year} {2016}{\natexlab{a}})},\ \Eprint
  {http://arxiv.org/abs/1606.06097} {arXiv:1606.06097} \BibitemShut {NoStop}%
\bibitem [{\citenamefont {Evans}\ \emph {et~al.}(2006)\citenamefont {Evans},
  \citenamefont {Thiel},\ and\ \citenamefont {Bartelt}}]{evans2006}%
  \BibitemOpen
  \bibfield  {author} {\bibinfo {author} {\bibfnamefont {J.}~\bibnamefont
  {Evans}}, \bibinfo {author} {\bibfnamefont {P.}~\bibnamefont {Thiel}}, \ and\
  \bibinfo {author} {\bibfnamefont {M.~C.}\ \bibnamefont {Bartelt}},\
  }\bibfield  {title} {\enquote {\bibinfo {title} {Morphological evolution
  during epitaxial thin film growth: {F}ormation of {2D} islands and {3D}
  mounds},}\ }\href {http://dx.doi.org/10.1016/j.surfrep.2005.08.004}
  {\bibfield  {journal} {\bibinfo  {journal} {Surf. Sci. Rep.}\ }\textbf
  {\bibinfo {volume} {61}},\ \bibinfo {pages} {1} (\bibinfo {year}
  {2006})}\BibitemShut {NoStop}%
\bibitem [{\citenamefont {Michely}\ and\ \citenamefont
  {Krug}(2004)}]{michely2004islands}%
  \BibitemOpen
  \bibfield  {author} {\bibinfo {author} {\bibfnamefont {T.}~\bibnamefont
  {Michely}}\ and\ \bibinfo {author} {\bibfnamefont {J.}~\bibnamefont {Krug}},\
  }\href {http://www.springer.com/us/book/9783540407287} {\emph {\bibinfo
  {title} {Islands, Mounds and Atoms}}}\ (\bibinfo  {publisher} {Springer},\
  \bibinfo {address} {Berlin},\ \bibinfo {year} {2004})\BibitemShut {NoStop}%
\bibitem [{\citenamefont {Meakin}(1998)}]{meakin1998fractals}%
  \BibitemOpen
  \bibfield  {author} {\bibinfo {author} {\bibfnamefont {P.}~\bibnamefont
  {Meakin}},\ }\href {http://books.google.com.br/books?id=cWmbNYSQDKoC} {\emph
  {\bibinfo {title} {Fractals, scaling and growth far from equilibrium}}}\
  (\bibinfo  {publisher} {Cambridge university press},\ \bibinfo {address}
  {Cambridge},\ \bibinfo {year} {1998})\BibitemShut {NoStop}%
\bibitem [{\citenamefont {Krim}\ and\ \citenamefont
  {Palasantzas}(1995)}]{Krim95}%
  \BibitemOpen
  \bibfield  {author} {\bibinfo {author} {\bibfnamefont {J.}~\bibnamefont
  {Krim}}\ and\ \bibinfo {author} {\bibfnamefont {G.}~\bibnamefont
  {Palasantzas}},\ }\bibfield  {title} {\enquote {\bibinfo {title}
  {Experimental observations of self-affine scaling and kinetic roughening at
  sub-micron lengthscales},}\ }\href
  {http://dx.doi.org/10.1142/S0217979295000239} {\bibfield  {journal} {\bibinfo
   {journal} {Int. J. Modern Phys. B}\ }\textbf {\bibinfo {volume} {09}},\
  \bibinfo {pages} {599} (\bibinfo {year} {1995})}\BibitemShut {NoStop}%
\bibitem [{\citenamefont {Palasantzas}\ \emph {et~al.}(2002)\citenamefont
  {Palasantzas}, \citenamefont {Tsamouras},\ and\ \citenamefont
  {Hosson}}]{Palasantzas02}%
  \BibitemOpen
  \bibfield  {author} {\bibinfo {author} {\bibfnamefont {G.}~\bibnamefont
  {Palasantzas}}, \bibinfo {author} {\bibfnamefont {D.}~\bibnamefont
  {Tsamouras}}, \ and\ \bibinfo {author} {\bibfnamefont {J.~D.}\ \bibnamefont
  {Hosson}},\ }\bibfield  {title} {\enquote {\bibinfo {title} {Roughening
  aspects of room temperature vapor deposited oligomer thin films onto {Si}
  substrates},}\ }\href {http://dx.doi.org/10.1016/S0039-6028(02)01271-2}
  {\bibfield  {journal} {\bibinfo  {journal} {Surf. Sci.}\ }\textbf {\bibinfo
  {volume} {507}},\ \bibinfo {pages} {357 } (\bibinfo {year}
  {2002})}\BibitemShut {NoStop}%
\bibitem [{\citenamefont {Ojeda}\ \emph {et~al.}(2000)\citenamefont {Ojeda},
  \citenamefont {Cuerno}, \citenamefont {Salvarezza},\ and\ \citenamefont
  {V\'azquez}}]{Ojeda2000}%
  \BibitemOpen
  \bibfield  {author} {\bibinfo {author} {\bibfnamefont {F.}~\bibnamefont
  {Ojeda}}, \bibinfo {author} {\bibfnamefont {R.}~\bibnamefont {Cuerno}},
  \bibinfo {author} {\bibfnamefont {R.}~\bibnamefont {Salvarezza}}, \ and\
  \bibinfo {author} {\bibfnamefont {L.}~\bibnamefont {V\'azquez}},\ }\bibfield
  {title} {\enquote {\bibinfo {title} {Dynamics of rough interfaces in chemical
  vapor deposition: Experiments and a model for silica films},}\ }\href
  {http://dx.doi.org/10.1103/PhysRevLett.84.3125} {\bibfield  {journal}
  {\bibinfo  {journal} {Phys. Rev. Lett.}\ }\textbf {\bibinfo {volume} {84}},\
  \bibinfo {pages} {3125} (\bibinfo {year} {2000})}\BibitemShut {NoStop}%
\bibitem [{\citenamefont {Kelling}\ and\ \citenamefont
  {\'Odor}(2011)}]{Kelling}%
  \BibitemOpen
  \bibfield  {author} {\bibinfo {author} {\bibfnamefont {J.}~\bibnamefont
  {Kelling}}\ and\ \bibinfo {author} {\bibfnamefont {G.}~\bibnamefont
  {\'Odor}},\ }\bibfield  {title} {\enquote {\bibinfo {title} {Extremely
  large-scale simulation of a {Kardar-Parisi-Zhang} model using graphics
  cards},}\ }\href {http://dx.doi.org/10.1103/PhysRevE.84.061150} {\bibfield
  {journal} {\bibinfo  {journal} {Phys. Rev. E}\ }\textbf {\bibinfo {volume}
  {84}},\ \bibinfo {pages} {061150} (\bibinfo {year} {2011})}\BibitemShut
  {NoStop}%
\bibitem [{\citenamefont {Halpin-Healy}\ and\ \citenamefont
  {Palasantzas}(2014)}]{MF2014}%
  \BibitemOpen
  \bibfield  {author} {\bibinfo {author} {\bibfnamefont {T.}~\bibnamefont
  {Halpin-Healy}}\ and\ \bibinfo {author} {\bibfnamefont {G.}~\bibnamefont
  {Palasantzas}},\ }\bibfield  {title} {\enquote {\bibinfo {title} {Universal
  correlators and distributions as experimental signatures of (2+1)-dimensional
  {Kardar-Parisi-Zhang growth}},}\ }\href@noop {} {\bibfield  {journal}
  {\bibinfo  {journal} {Europhys. Lett.}\ }\textbf {\bibinfo {volume} {105}},\
  \bibinfo {pages} {50001} (\bibinfo {year} {2014})}\BibitemShut {NoStop}%
\bibitem [{\citenamefont {Almeida}\ \emph {et~al.}(2014)\citenamefont
  {Almeida}, \citenamefont {Ferreira}, \citenamefont {Oliveira},\ and\
  \citenamefont {Reis}}]{almeida2013}%
  \BibitemOpen
  \bibfield  {author} {\bibinfo {author} {\bibfnamefont {R.~A.~L.}\
  \bibnamefont {Almeida}}, \bibinfo {author} {\bibfnamefont {S.~O.}\
  \bibnamefont {Ferreira}}, \bibinfo {author} {\bibfnamefont {T.~J.}\
  \bibnamefont {Oliveira}}, \ and\ \bibinfo {author} {\bibfnamefont {F.~D.
  A.~A.}\ \bibnamefont {Reis}},\ }\bibfield  {title} {\enquote {\bibinfo
  {title} {Universal fluctuations in the growth of semiconductor thin films},}\
  }\href {http://dx.doi.org/10.1103/PhysRevB.89.045309} {\bibfield  {journal}
  {\bibinfo  {journal} {Phys. Rev. B}\ }\textbf {\bibinfo {volume} {89}},\
  \bibinfo {pages} {045309} (\bibinfo {year} {2014})}\BibitemShut {NoStop}%
\bibitem [{\citenamefont {Almeida}\ \emph {et~al.}(2015)\citenamefont
  {Almeida}, \citenamefont {Ferreira}, \citenamefont {Ribeiro},\ and\
  \citenamefont {Oliveira}}]{Almeida2015}%
  \BibitemOpen
  \bibfield  {author} {\bibinfo {author} {\bibfnamefont {R.~A.~L.}\
  \bibnamefont {Almeida}}, \bibinfo {author} {\bibfnamefont {S.~O.}\
  \bibnamefont {Ferreira}}, \bibinfo {author} {\bibfnamefont {I.~R.~B.}\
  \bibnamefont {Ribeiro}}, \ and\ \bibinfo {author} {\bibfnamefont {T.~J.}\
  \bibnamefont {Oliveira}},\ }\bibfield  {title} {\enquote {\bibinfo {title}
  {Temperature effect on (2 + 1) experimental {Kardar-Parisi-Zhang} growth},}\
  }\href {http://dx.doi.org/10.1209/0295-5075/109/46003} {\bibfield  {journal}
  {\bibinfo  {journal} {Europhys. Lett.}\ }\textbf {\bibinfo {volume} {109}},\
  \bibinfo {pages} {46003} (\bibinfo {year} {2015})}\BibitemShut {NoStop}%
\bibitem [{\citenamefont {R\'acz}\ and\ \citenamefont {Plischke}(1994)}]{Racz}%
  \BibitemOpen
  \bibfield  {author} {\bibinfo {author} {\bibfnamefont {Z.}~\bibnamefont
  {R\'acz}}\ and\ \bibinfo {author} {\bibfnamefont {M.}~\bibnamefont
  {Plischke}},\ }\bibfield  {title} {\enquote {\bibinfo {title} {Width
  distribution for (2+1)-dimensional growth and deposition processes},}\ }\href
  {http://dx.doi.org/10.1103/PhysRevE.50.3530} {\bibfield  {journal} {\bibinfo
  {journal} {Phys. Rev. E}\ }\textbf {\bibinfo {volume} {50}},\ \bibinfo
  {pages} {3530} (\bibinfo {year} {1994})}\BibitemShut {NoStop}%
\bibitem [{\citenamefont {Raychaudhuri}\ \emph {et~al.}(2001)\citenamefont
  {Raychaudhuri}, \citenamefont {Cranston}, \citenamefont {Przybyla},\ and\
  \citenamefont {Shapir}}]{Raychaudhuri}%
  \BibitemOpen
  \bibfield  {author} {\bibinfo {author} {\bibfnamefont {S.}~\bibnamefont
  {Raychaudhuri}}, \bibinfo {author} {\bibfnamefont {M.}~\bibnamefont
  {Cranston}}, \bibinfo {author} {\bibfnamefont {C.}~\bibnamefont {Przybyla}},
  \ and\ \bibinfo {author} {\bibfnamefont {Y.}~\bibnamefont {Shapir}},\
  }\bibfield  {title} {\enquote {\bibinfo {title} {Maximal height scaling of
  kinetically growing surfaces},}\ }\href
  {http://dx.doi.org/10.1103/PhysRevLett.87.136101} {\bibfield  {journal}
  {\bibinfo  {journal} {Phys. Rev. Lett.}\ }\textbf {\bibinfo {volume} {87}},\
  \bibinfo {pages} {136101} (\bibinfo {year} {2001})}\BibitemShut {NoStop}%
\bibitem [{\citenamefont {Majumdar}\ and\ \citenamefont
  {Comtet}(2004)}]{Majumdar04}%
  \BibitemOpen
  \bibfield  {author} {\bibinfo {author} {\bibfnamefont {S.~N.}\ \bibnamefont
  {Majumdar}}\ and\ \bibinfo {author} {\bibfnamefont {A.}~\bibnamefont
  {Comtet}},\ }\bibfield  {title} {\enquote {\bibinfo {title} {Exact maximal
  height distribution of fluctuating interfaces},}\ }\href
  {http://dx.doi.org/10.1103/PhysRevLett.92.225501} {\bibfield  {journal}
  {\bibinfo  {journal} {Phys. Rev. Lett.}\ }\textbf {\bibinfo {volume} {92}},\
  \bibinfo {pages} {225501} (\bibinfo {year} {2004})}\BibitemShut {NoStop}%
\bibitem [{\citenamefont {Lechenault}\ \emph {et~al.}(2010)\citenamefont
  {Lechenault}, \citenamefont {Pallares}, \citenamefont {George}, \citenamefont
  {Rountree}, \citenamefont {Bouchaud},\ and\ \citenamefont
  {Ciccotti}}]{Lechenault}%
  \BibitemOpen
  \bibfield  {author} {\bibinfo {author} {\bibfnamefont {F.}~\bibnamefont
  {Lechenault}}, \bibinfo {author} {\bibfnamefont {G.}~\bibnamefont
  {Pallares}}, \bibinfo {author} {\bibfnamefont {M.}~\bibnamefont {George}},
  \bibinfo {author} {\bibfnamefont {C.}~\bibnamefont {Rountree}}, \bibinfo
  {author} {\bibfnamefont {E.}~\bibnamefont {Bouchaud}}, \ and\ \bibinfo
  {author} {\bibfnamefont {M.}~\bibnamefont {Ciccotti}},\ }\bibfield  {title}
  {\enquote {\bibinfo {title} {Effects of finite probe size on self-affine
  roughness measurements},}\ }\href
  {http://dx.doi.org/10.1103/PhysRevLett.104.025502} {\bibfield  {journal}
  {\bibinfo  {journal} {Phys. Rev. Lett.}\ }\textbf {\bibinfo {volume} {104}},\
  \bibinfo {pages} {025502} (\bibinfo {year} {2010})}\BibitemShut {NoStop}%
\bibitem [{\citenamefont {Villarrubia}(1994)}]{villarrubia94}%
  \BibitemOpen
  \bibfield  {author} {\bibinfo {author} {\bibfnamefont {J.}~\bibnamefont
  {Villarrubia}},\ }\bibfield  {title} {\enquote {\bibinfo {title}
  {Morphological estimation of tip geometry for scanned probe microscopy},}\
  }\href {http://dx.doi.org/10.1016/0039-6028(94)90194-5} {\bibfield  {journal}
  {\bibinfo  {journal} {Surface Science}\ }\textbf {\bibinfo {volume} {321}},\
  \bibinfo {pages} {287 } (\bibinfo {year} {1994})}\BibitemShut {NoStop}%
\bibitem [{\citenamefont {Wolf}\ and\ \citenamefont {Villain}(1990)}]{WV}%
  \BibitemOpen
  \bibfield  {author} {\bibinfo {author} {\bibfnamefont {D.~E.}\ \bibnamefont
  {Wolf}}\ and\ \bibinfo {author} {\bibfnamefont {J.}~\bibnamefont {Villain}},\
  }\bibfield  {title} {\enquote {\bibinfo {title} {Growth with surface
  diffusion},}\ }\href {http://dx.doi.org/10.1209/0295-5075/13/5/002}
  {\bibfield  {journal} {\bibinfo  {journal} {Europhys. Lett.}\ }\textbf
  {\bibinfo {volume} {13}},\ \bibinfo {pages} {389} (\bibinfo {year}
  {1990})}\BibitemShut {NoStop}%
\bibitem [{\citenamefont {Das~Sarma}\ and\ \citenamefont
  {Tamborenea}(1991)}]{DT}%
  \BibitemOpen
  \bibfield  {author} {\bibinfo {author} {\bibfnamefont {S.}~\bibnamefont
  {Das~Sarma}}\ and\ \bibinfo {author} {\bibfnamefont {P.}~\bibnamefont
  {Tamborenea}},\ }\bibfield  {title} {\enquote {\bibinfo {title} {A new
  universality class for kinetic growth: One-dimensional molecular-beam
  epitaxy},}\ }\href {http://dx.doi.org/10.1103/PhysRevLett.66.325} {\bibfield
  {journal} {\bibinfo  {journal} {Phys. Rev. Lett.}\ }\textbf {\bibinfo
  {volume} {66}},\ \bibinfo {pages} {325} (\bibinfo {year} {1991})}\BibitemShut
  {NoStop}%
\bibitem [{\citenamefont {Chatraphorn}\ \emph {et~al.}(2001)\citenamefont
  {Chatraphorn}, \citenamefont {Toroczkai},\ and\ \citenamefont
  {Das~Sarma}}]{Chatraphorn}%
  \BibitemOpen
  \bibfield  {author} {\bibinfo {author} {\bibfnamefont {P.~P.}\ \bibnamefont
  {Chatraphorn}}, \bibinfo {author} {\bibfnamefont {Z.}~\bibnamefont
  {Toroczkai}}, \ and\ \bibinfo {author} {\bibfnamefont {S.}~\bibnamefont
  {Das~Sarma}},\ }\bibfield  {title} {\enquote {\bibinfo {title} {Epitaxial
  mounding in limited-mobility models of surface growth},}\ }\href {\doibase
  10.1103/PhysRevB.64.205407} {\bibfield  {journal} {\bibinfo  {journal} {Phys.
  Rev. B}\ }\textbf {\bibinfo {volume} {64}},\ \bibinfo {pages} {205407}
  (\bibinfo {year} {2001})}\BibitemShut {NoStop}%
\bibitem [{\citenamefont {Alves}\ \emph {et~al.}(2011)\citenamefont {Alves},
  \citenamefont {Oliveira},\ and\ \citenamefont {Ferreira}}]{Alves11}%
  \BibitemOpen
  \bibfield  {author} {\bibinfo {author} {\bibfnamefont {S.~G.}\ \bibnamefont
  {Alves}}, \bibinfo {author} {\bibfnamefont {T.~J.}\ \bibnamefont {Oliveira}},
  \ and\ \bibinfo {author} {\bibfnamefont {S.~C.}\ \bibnamefont {Ferreira}},\
  }\bibfield  {title} {\enquote {\bibinfo {title} {Universal fluctuations in
  radial growth models belonging to the {KPZ} universality class},}\ }\href
  {http://dx.doi.org/10.1209/0295-5075/96/48003} {\bibfield  {journal}
  {\bibinfo  {journal} {Europhys. Lett.}\ }\textbf {\bibinfo {volume} {96}},\
  \bibinfo {pages} {48003} (\bibinfo {year} {2011})}\BibitemShut {NoStop}%
\bibitem [{\citenamefont {Kim}\ and\ \citenamefont {Kosterlitz}(1989)}]{kk}%
  \BibitemOpen
  \bibfield  {author} {\bibinfo {author} {\bibfnamefont {J.~M.}\ \bibnamefont
  {Kim}}\ and\ \bibinfo {author} {\bibfnamefont {J.~M.}\ \bibnamefont
  {Kosterlitz}},\ }\bibfield  {title} {\enquote {\bibinfo {title} {Growth in a
  restricted solid-on-solid model},}\ }\href
  {http://dx.doi.org/10.1103/PhysRevLett.62.2289} {\bibfield  {journal}
  {\bibinfo  {journal} {Phys. Rev. Lett.}\ }\textbf {\bibinfo {volume} {62}},\
  \bibinfo {pages} {2289} (\bibinfo {year} {1989})}\BibitemShut {NoStop}%
\bibitem [{\citenamefont {Alves}\ \emph
  {et~al.}(2014{\natexlab{a}})\citenamefont {Alves}, \citenamefont {Oliveira},\
  and\ \citenamefont {Ferreira}}]{Alves14}%
  \BibitemOpen
  \bibfield  {author} {\bibinfo {author} {\bibfnamefont {S.~G.}\ \bibnamefont
  {Alves}}, \bibinfo {author} {\bibfnamefont {T.~J.}\ \bibnamefont {Oliveira}},
  \ and\ \bibinfo {author} {\bibfnamefont {S.~C.}\ \bibnamefont {Ferreira}},\
  }\bibfield  {title} {\enquote {\bibinfo {title} {Universality of fluctuations
  in the {Kardar-Parisi-Zhang} class in high dimensions and its upper critical
  dimension},}\ }\href {http://dx.doi.org/10.1103/PhysRevE.90.020103}
  {\bibfield  {journal} {\bibinfo  {journal} {Phys. Rev. E}\ }\textbf {\bibinfo
  {volume} {90}},\ \bibinfo {pages} {020103} (\bibinfo {year}
  {2014}{\natexlab{a}})}\BibitemShut {NoStop}%
\bibitem [{\citenamefont {Yunker}\ \emph {et~al.}(2013)\citenamefont {Yunker},
  \citenamefont {Lohr}, \citenamefont {Still}, \citenamefont {Borodin},
  \citenamefont {Durian},\ and\ \citenamefont {Yodh}}]{yunker}%
  \BibitemOpen
  \bibfield  {author} {\bibinfo {author} {\bibfnamefont {P.~J.}\ \bibnamefont
  {Yunker}}, \bibinfo {author} {\bibfnamefont {M.~A.}\ \bibnamefont {Lohr}},
  \bibinfo {author} {\bibfnamefont {T.}~\bibnamefont {Still}}, \bibinfo
  {author} {\bibfnamefont {A.}~\bibnamefont {Borodin}}, \bibinfo {author}
  {\bibfnamefont {D.~J.}\ \bibnamefont {Durian}}, \ and\ \bibinfo {author}
  {\bibfnamefont {A.~G.}\ \bibnamefont {Yodh}},\ }\bibfield  {title} {\enquote
  {\bibinfo {title} {Effects of particle shape on growth dynamics at edges of
  evaporating drops of colloidal suspensions},}\ }\href
  {http://dx.doi.org/10.1103/PhysRevLett.110.035501} {\bibfield  {journal}
  {\bibinfo  {journal} {Phys. Rev. Lett.}\ }\textbf {\bibinfo {volume} {110}},\
  \bibinfo {pages} {035501} (\bibinfo {year} {2013})}\BibitemShut {NoStop}%
\bibitem [{\citenamefont {de~Castro}\ \emph {et~al.}(2014)\citenamefont
  {de~Castro}, \citenamefont {de~Assis}, \citenamefont {de~Castilho},\ and\
  \citenamefont {Andrade}}]{Castro14}%
  \BibitemOpen
  \bibfield  {author} {\bibinfo {author} {\bibfnamefont {C.~P.}\ \bibnamefont
  {de~Castro}}, \bibinfo {author} {\bibfnamefont {T.~A.}\ \bibnamefont
  {de~Assis}}, \bibinfo {author} {\bibfnamefont {C.~M.~C.}\ \bibnamefont
  {de~Castilho}}, \ and\ \bibinfo {author} {\bibfnamefont {R.~F.~S.}\
  \bibnamefont {Andrade}},\ }\bibfield  {title} {\enquote {\bibinfo {title}
  {Height distribution of equipotential lines in a region confined by a rough
  conducting boundary},}\ }\href
  {http://dx.doi.org/10.1088/0953-8984/26/44/445007} {\bibfield  {journal}
  {\bibinfo  {journal} {J. Phys.: Cond. Matter}\ }\textbf {\bibinfo {volume}
  {26}},\ \bibinfo {pages} {445007} (\bibinfo {year} {2014})}\BibitemShut
  {NoStop}%
\bibitem [{\citenamefont {Carrasco}\ and\ \citenamefont
  {Oliveira}(2016{\natexlab{b}})}]{Carrasco16}%
  \BibitemOpen
  \bibfield  {author} {\bibinfo {author} {\bibfnamefont {I.~S.~S.}\
  \bibnamefont {Carrasco}}\ and\ \bibinfo {author} {\bibfnamefont {T.~J.}\
  \bibnamefont {Oliveira}},\ }\bibfield  {title} {\enquote {\bibinfo {title}
  {Width and extremal height distributions of fluctuating interfaces with
  window boundary conditions},}\ }\href
  {http://link.aps.org/doi/10.1103/PhysRevE.93.012801} {\bibfield  {journal}
  {\bibinfo  {journal} {Phys. Rev. E}\ }\textbf {\bibinfo {volume} {93}},\
  \bibinfo {pages} {012801} (\bibinfo {year} {2016}{\natexlab{b}})}\BibitemShut
  {NoStop}%
\bibitem [{\citenamefont {{Rambeau, J.}}\ and\ \citenamefont {{Schehr,
  G.}}(2010)}]{Rambeau10}%
  \BibitemOpen
  \bibfield  {author} {\bibinfo {author} {\bibnamefont {{Rambeau, J.}}}\ and\
  \bibinfo {author} {\bibnamefont {{Schehr, G.}}},\ }\bibfield  {title}
  {\enquote {\bibinfo {title} {Extremal statistics of curved growing interfaces
  in 1+1 dimensions},}\ }\href {http://dx.doi.org/10.1209/0295-5075/91/60006}
  {\bibfield  {journal} {\bibinfo  {journal} {Europhys. lett.}\ }\textbf
  {\bibinfo {volume} {91}},\ \bibinfo {pages} {60006} (\bibinfo {year}
  {2010})}\BibitemShut {NoStop}%
\bibitem [{\citenamefont {Halpin-Healy}\ and\ \citenamefont
  {Takeuchi}(2015)}]{MF2015}%
  \BibitemOpen
  \bibfield  {author} {\bibinfo {author} {\bibfnamefont {T.}~\bibnamefont
  {Halpin-Healy}}\ and\ \bibinfo {author} {\bibfnamefont {K.~A.}\ \bibnamefont
  {Takeuchi}},\ }\bibfield  {title} {\enquote {\bibinfo {title} {A {KPZ}
  cocktail-shaken, not stirred...}}\ }\href@noop {} {\bibfield  {journal}
  {\bibinfo  {journal} {J. Stat. Phys.}\ }\textbf {\bibinfo {volume} {160}},\
  \bibinfo {pages} {794} (\bibinfo {year} {2015})}\BibitemShut {NoStop}%
\bibitem [{\citenamefont {Reis}(2015)}]{Fabio2015}%
  \BibitemOpen
  \bibfield  {author} {\bibinfo {author} {\bibfnamefont {F.~D. A.~A.}\
  \bibnamefont {Reis}},\ }\bibfield  {title} {\enquote {\bibinfo {title}
  {Scaling of local roughness distributions},}\ }\href
  {http://stacks.iop.org/1742-5468/2015/i=11/a=P11020} {\bibfield  {journal}
  {\bibinfo  {journal} {J. Stat. Mech.}\ }\textbf {\bibinfo {volume} {2015}},\
  \bibinfo {pages} {P11020} (\bibinfo {year} {2015})}\BibitemShut {NoStop}%
\bibitem [{\citenamefont {Takeuchi}\ and\ \citenamefont
  {Sano}(2012)}]{TakeuchiJSP}%
  \BibitemOpen
  \bibfield  {author} {\bibinfo {author} {\bibfnamefont {K.}~\bibnamefont
  {Takeuchi}}\ and\ \bibinfo {author} {\bibfnamefont {M.}~\bibnamefont
  {Sano}},\ }\bibfield  {title} {\enquote {\bibinfo {title} {Evidence for
  geometry-dependent universal fluctuations of the {K}ardar-{P}arisi-{Z}hang
  interfaces in liquid-crystal turbulence},}\ }\href
  {http://link.springer.com/article/10.1007/s10955-012-0503-0} {\bibfield
  {journal} {\bibinfo  {journal} {J. Stat. Phys.}\ }\textbf {\bibinfo {volume}
  {147}},\ \bibinfo {pages} {853} (\bibinfo {year} {2012})}\BibitemShut
  {NoStop}%
\bibitem [{\citenamefont {Alves}\ \emph {et~al.}(2013)\citenamefont {Alves},
  \citenamefont {Oliveira},\ and\ \citenamefont {Ferreira}}]{Alves13}%
  \BibitemOpen
  \bibfield  {author} {\bibinfo {author} {\bibfnamefont {S.~G.}\ \bibnamefont
  {Alves}}, \bibinfo {author} {\bibfnamefont {T.~J.}\ \bibnamefont {Oliveira}},
  \ and\ \bibinfo {author} {\bibfnamefont {S.~C.}\ \bibnamefont {Ferreira}},\
  }\bibfield  {title} {\enquote {\bibinfo {title} {Non-universal parameters,
  corrections and universality in {Kardar-Parisi-Zhang} growth},}\ }\href
  {http://dx.doi.org/10.1088/1742-5468/2013/05/P05007} {\bibfield  {journal}
  {\bibinfo  {journal} {J. Stat. Mech.: Theor. Exp.}\ }\textbf {\bibinfo
  {volume} {2013}},\ \bibinfo {pages} {P05007} (\bibinfo {year}
  {2013})}\BibitemShut {NoStop}%
\bibitem [{\citenamefont {Alves}\ and\ \citenamefont
  {Moreira}(2011)}]{AlvesMoreira}%
  \BibitemOpen
  \bibfield  {author} {\bibinfo {author} {\bibfnamefont {S.~G.}\ \bibnamefont
  {Alves}}\ and\ \bibinfo {author} {\bibfnamefont {J.~G.}\ \bibnamefont
  {Moreira}},\ }\bibfield  {title} {\enquote {\bibinfo {title} {Transitions in
  a probabilistic interface growth model},}\ }\href
  {http://stacks.iop.org/1742-5468/2011/i=04/a=P04022} {\bibfield  {journal}
  {\bibinfo  {journal} {J. Stat. Mech.}\ }\textbf {\bibinfo {volume} {2011}},\
  \bibinfo {pages} {P04022} (\bibinfo {year} {2011})}\BibitemShut {NoStop}%
\bibitem [{\citenamefont {Leal}\ \emph {et~al.}(2011)\citenamefont {Leal},
  \citenamefont {Oliveira},\ and\ \citenamefont {Ferreira}}]{Leal11a}%
  \BibitemOpen
  \bibfield  {author} {\bibinfo {author} {\bibfnamefont {F.~F.}\ \bibnamefont
  {Leal}}, \bibinfo {author} {\bibfnamefont {T.~J.}\ \bibnamefont {Oliveira}},
  \ and\ \bibinfo {author} {\bibfnamefont {S.~C.}\ \bibnamefont {Ferreira}},\
  }\bibfield  {title} {\enquote {\bibinfo {title} {Kinetic modelling of
  epitaxial film growth with {up- and} downward step barriers},}\ }\href
  {http://stacks.iop.org/1742-5468/2011/i=09/a=P09018} {\bibfield  {journal}
  {\bibinfo  {journal} {J. Stat. Mech.}\ }\textbf {\bibinfo {volume} {2011}},\
  \bibinfo {pages} {P09018} (\bibinfo {year} {2011})}\BibitemShut {NoStop}%
\bibitem [{\citenamefont {Zorba}\ \emph {et~al.}(2006)\citenamefont {Zorba},
  \citenamefont {Shapir},\ and\ \citenamefont {Gao}}]{Zorba}%
  \BibitemOpen
  \bibfield  {author} {\bibinfo {author} {\bibfnamefont {S.}~\bibnamefont
  {Zorba}}, \bibinfo {author} {\bibfnamefont {Y.}~\bibnamefont {Shapir}}, \
  and\ \bibinfo {author} {\bibfnamefont {Y.}~\bibnamefont {Gao}},\ }\bibfield
  {title} {\enquote {\bibinfo {title} {Fractal-mound growth of pentacene thin
  films},}\ }\href {http://link.aps.org/doi/10.1103/PhysRevB.74.245410}
  {\bibfield  {journal} {\bibinfo  {journal} {Phys. Rev. B}\ }\textbf {\bibinfo
  {volume} {74}},\ \bibinfo {pages} {245410} (\bibinfo {year}
  {2006})}\BibitemShut {NoStop}%
\bibitem [{\citenamefont {Hlawacek}\ \emph {et~al.}(2008)\citenamefont
  {Hlawacek}, \citenamefont {Puschnig}, \citenamefont {Frank}, \citenamefont
  {Winkler}, \citenamefont {Ambrosch-Draxl},\ and\ \citenamefont
  {Teichert}}]{Hlawacek108}%
  \BibitemOpen
  \bibfield  {author} {\bibinfo {author} {\bibfnamefont {G.}~\bibnamefont
  {Hlawacek}}, \bibinfo {author} {\bibfnamefont {P.}~\bibnamefont {Puschnig}},
  \bibinfo {author} {\bibfnamefont {P.}~\bibnamefont {Frank}}, \bibinfo
  {author} {\bibfnamefont {A.}~\bibnamefont {Winkler}}, \bibinfo {author}
  {\bibfnamefont {C.}~\bibnamefont {Ambrosch-Draxl}}, \ and\ \bibinfo {author}
  {\bibfnamefont {C.}~\bibnamefont {Teichert}},\ }\bibfield  {title} {\enquote
  {\bibinfo {title} {Characterization of step-edge barriers in organic
  thin-film growth},}\ }\href {http://dx.doi.org/10.1126/science.1159455}
  {\bibfield  {journal} {\bibinfo  {journal} {Science}\ }\textbf {\bibinfo
  {volume} {321}},\ \bibinfo {pages} {108} (\bibinfo {year}
  {2008})}\BibitemShut {NoStop}%
\bibitem [{\citenamefont {Family}(1986)}]{Family}%
  \BibitemOpen
  \bibfield  {author} {\bibinfo {author} {\bibfnamefont {F.}~\bibnamefont
  {Family}},\ }\bibfield  {title} {\enquote {\bibinfo {title} {Scaling of rough
  surfaces: effects of surface diffusion},}\ }\href
  {http://dx.doi.org/10.1088/0305-4470/19/8/006} {\bibfield  {journal}
  {\bibinfo  {journal} {J. Phys. A: Math.Gen.}\ }\textbf {\bibinfo {volume}
  {19}},\ \bibinfo {pages} {L441} (\bibinfo {year} {1986})}\BibitemShut
  {NoStop}%
\bibitem [{\citenamefont {Sagu\'{e}s}\ \emph {et~al.}(2000)\citenamefont
  {Sagu\'{e}s}, \citenamefont {L\'{o}pez-Salvans},\ and\ \citenamefont
  {Claret}}]{Sagues2000}%
  \BibitemOpen
  \bibfield  {author} {\bibinfo {author} {\bibfnamefont {F.}~\bibnamefont
  {Sagu\'{e}s}}, \bibinfo {author} {\bibfnamefont {M.~Q.}\ \bibnamefont
  {L\'{o}pez-Salvans}}, \ and\ \bibinfo {author} {\bibfnamefont
  {J.}~\bibnamefont {Claret}},\ }\bibfield  {title} {\enquote {\bibinfo {title}
  {Growth and forms in quasi-two-dimensional electrocrystallization},}\ }\href
  {http://dx.doi.org/10.1016/S0370-1573(00)00057-0} {\bibfield  {journal}
  {\bibinfo  {journal} {Phys. Rep.}\ }\textbf {\bibinfo {volume} {337}},\
  \bibinfo {pages} {97 } (\bibinfo {year} {2000})}\BibitemShut {NoStop}%
\bibitem [{\citenamefont {Lafouresse}\ \emph {et~al.}(2007)\citenamefont
  {Lafouresse}, \citenamefont {Heard},\ and\ \citenamefont
  {Schwarzacher}}]{lafouresse}%
  \BibitemOpen
  \bibfield  {author} {\bibinfo {author} {\bibfnamefont {M.}~\bibnamefont
  {Lafouresse}}, \bibinfo {author} {\bibfnamefont {P.}~\bibnamefont {Heard}}, \
  and\ \bibinfo {author} {\bibfnamefont {W.}~\bibnamefont {Schwarzacher}},\
  }\bibfield  {title} {\enquote {\bibinfo {title} {Anomalous scaling for thick
  electrodeposited films},}\ }\href
  {http://dx.doi.org/10.1103/PhysRevLett.98.236101} {\bibfield  {journal}
  {\bibinfo  {journal} {Phys. Rev. Lett.}\ }\textbf {\bibinfo {volume} {98}},\
  \bibinfo {pages} {236101} (\bibinfo {year} {2007})}\BibitemShut {NoStop}%
\bibitem [{\citenamefont {C\'ordoba-Torres}\ \emph {et~al.}(2009)\citenamefont
  {C\'ordoba-Torres}, \citenamefont {Mesquita}, \citenamefont {Bastos},\ and\
  \citenamefont {Nogueira}}]{Cordoba-Torres}%
  \BibitemOpen
  \bibfield  {author} {\bibinfo {author} {\bibfnamefont {P.}~\bibnamefont
  {C\'ordoba-Torres}}, \bibinfo {author} {\bibfnamefont {T.~J.}\ \bibnamefont
  {Mesquita}}, \bibinfo {author} {\bibfnamefont {I.~N.}\ \bibnamefont
  {Bastos}}, \ and\ \bibinfo {author} {\bibfnamefont {R.~P.}\ \bibnamefont
  {Nogueira}},\ }\bibfield  {title} {\enquote {\bibinfo {title} {Complex
  dynamics during metal dissolution: From intrinsic to faceted anomalous
  scaling},}\ }\href {http://dx.doi.org/10.1103/PhysRevLett.102.055504}
  {\bibfield  {journal} {\bibinfo  {journal} {Phys. Rev. Lett.}\ }\textbf
  {\bibinfo {volume} {102}},\ \bibinfo {pages} {055504} (\bibinfo {year}
  {2009})}\BibitemShut {NoStop}%
\bibitem [{\citenamefont {Felix}\ \emph {et~al.}(2015)\citenamefont {Felix},
  \citenamefont {Figueiredo}, \citenamefont {Mendes}, \citenamefont {Morais},\
  and\ \citenamefont {de. Araujo}}]{cloclo}%
  \BibitemOpen
  \bibfield  {author} {\bibinfo {author} {\bibfnamefont {J.}~\bibnamefont
  {Felix}}, \bibinfo {author} {\bibfnamefont {L.}~\bibnamefont {Figueiredo}},
  \bibinfo {author} {\bibfnamefont {J.}~\bibnamefont {Mendes}}, \bibinfo
  {author} {\bibfnamefont {P.}~\bibnamefont {Morais}}, \ and\ \bibinfo {author}
  {\bibfnamefont {C.}~\bibnamefont {de. Araujo}},\ }\bibfield  {title}
  {\enquote {\bibinfo {title} {Low-field microwave absorption and
  magnetoresistance in iron nanostructures grown by electrodeposition on n-type
  lightly doped silicon substrates},}\ }\href {\doibase
  http://dx.doi.org/10.1016/j.jmmm.2015.07.061} {\bibfield  {journal} {\bibinfo
   {journal} {J. Magn. Magn. Mater.}\ }\textbf {\bibinfo {volume} {395}},\
  \bibinfo {pages} {130 } (\bibinfo {year} {2015})}\BibitemShut {NoStop}%
\bibitem [{\citenamefont {Guzm\'{a}n-Casta\~{n}eda}\ \emph
  {et~al.}(2012)\citenamefont {Guzm\'{a}n-Casta\~{n}eda}, \citenamefont
  {Garc\'{\i}ca-B\'{o}rquez},\ and\ \citenamefont
  {Arizabalo-Salas}}]{Castaneda}%
  \BibitemOpen
  \bibfield  {author} {\bibinfo {author} {\bibfnamefont {J.~I.}\ \bibnamefont
  {Guzm\'{a}n-Casta\~{n}eda}}, \bibinfo {author} {\bibfnamefont
  {A.}~\bibnamefont {Garc\'{\i}ca-B\'{o}rquez}}, \ and\ \bibinfo {author}
  {\bibfnamefont {R.~D.}\ \bibnamefont {Arizabalo-Salas}},\ }\bibfield  {title}
  {\enquote {\bibinfo {title} {Fractal dimension determined through optical and
  scanning electron microscopy on fecral alloy after polishing, erosion, and
  oxidizing processes},}\ }\href {http://dx.doi.org/10.1002/pssb.201100806}
  {\bibfield  {journal} {\bibinfo  {journal} {Physica Status Solidi (b)}\
  }\textbf {\bibinfo {volume} {249}},\ \bibinfo {pages} {1224} (\bibinfo {year}
  {2012})}\BibitemShut {NoStop}%
\bibitem [{\citenamefont {{\'O}dor}\ \emph {et~al.}(2014)\citenamefont
  {{\'O}dor}, \citenamefont {Kelling},\ and\ \citenamefont
  {Gemming}}]{odor2014aging}%
  \BibitemOpen
  \bibfield  {author} {\bibinfo {author} {\bibfnamefont {G.}~\bibnamefont
  {{\'O}dor}}, \bibinfo {author} {\bibfnamefont {J.}~\bibnamefont {Kelling}}, \
  and\ \bibinfo {author} {\bibfnamefont {S.}~\bibnamefont {Gemming}},\
  }\bibfield  {title} {\enquote {\bibinfo {title} {Aging of the (2+
  1)-dimensional {Kardar-Parisi-Zhang model}},}\ }\href
  {http://dx.doi.org/10.1103/PhysRevE.89.032146} {\bibfield  {journal}
  {\bibinfo  {journal} {Phys. Rev. E}\ }\textbf {\bibinfo {volume} {89}},\
  \bibinfo {pages} {032146} (\bibinfo {year} {2014})}\BibitemShut {NoStop}%
\bibitem [{\citenamefont {Carrasco}\ \emph {et~al.}(2014)\citenamefont
  {Carrasco}, \citenamefont {Takeuchi}, \citenamefont {Ferreira},\ and\
  \citenamefont {Oliveira}}]{Carrasco}%
  \BibitemOpen
  \bibfield  {author} {\bibinfo {author} {\bibfnamefont {I.~S.~S.}\
  \bibnamefont {Carrasco}}, \bibinfo {author} {\bibfnamefont {K.~A.}\
  \bibnamefont {Takeuchi}}, \bibinfo {author} {\bibfnamefont {S.~C.}\
  \bibnamefont {Ferreira}}, \ and\ \bibinfo {author} {\bibfnamefont {T.~J.}\
  \bibnamefont {Oliveira}},\ }\bibfield  {title} {\enquote {\bibinfo {title}
  {Interface fluctuations for deposition on enlarging flat substrates},}\
  }\href {http://dx.doi.org/10.1088/1367-2630/16/12/123057} {\bibfield
  {journal} {\bibinfo  {journal} {New J. Phys.}\ }\textbf {\bibinfo {volume}
  {16}},\ \bibinfo {pages} {123057} (\bibinfo {year} {2014})}\BibitemShut
  {NoStop}%
\bibitem [{\citenamefont {Krug}\ and\ \citenamefont {Meakin}(1990)}]{Krug90}%
  \BibitemOpen
  \bibfield  {author} {\bibinfo {author} {\bibfnamefont {J.}~\bibnamefont
  {Krug}}\ and\ \bibinfo {author} {\bibfnamefont {P.}~\bibnamefont {Meakin}},\
  }\bibfield  {title} {\enquote {\bibinfo {title} {Universal finite-size
  effects in the rate of growth processes},}\ }\href
  {http://dx.doi.org/10.1088/0305-4470/23/18/009} {\bibfield  {journal}
  {\bibinfo  {journal} {J. Phys. A: Math. Gen.}\ }\textbf {\bibinfo {volume}
  {23}},\ \bibinfo {pages} {L987} (\bibinfo {year} {1990})}\BibitemShut
  {NoStop}%
\bibitem [{\citenamefont {Takeuchi}(2014)}]{takeuchi2014}%
  \BibitemOpen
  \bibfield  {author} {\bibinfo {author} {\bibfnamefont {K.~A.}\ \bibnamefont
  {Takeuchi}},\ }\bibfield  {title} {\enquote {\bibinfo {title} {Experimental
  approaches to universal out-of-equilibrium scaling laws: turbulent liquid
  crystal and other developments},}\ }\href
  {http://stacks.iop.org/1742-5468/2014/i=1/a=P01006} {\bibfield  {journal}
  {\bibinfo  {journal} {J. Stat. Mech.}\ }\textbf {\bibinfo {volume} {2014}},\
  \bibinfo {pages} {P01006} (\bibinfo {year} {2014})}\BibitemShut {NoStop}%
\bibitem [{\citenamefont {Alves}\ \emph
  {et~al.}(2014{\natexlab{b}})\citenamefont {Alves}, \citenamefont {Oliveira},\
  and\ \citenamefont {Ferreira}}]{AlvesOrigins}%
  \BibitemOpen
  \bibfield  {author} {\bibinfo {author} {\bibfnamefont {S.~G.}\ \bibnamefont
  {Alves}}, \bibinfo {author} {\bibfnamefont {T.~J.}\ \bibnamefont {Oliveira}},
  \ and\ \bibinfo {author} {\bibfnamefont {S.~C.}\ \bibnamefont {Ferreira}},\
  }\bibfield  {title} {\enquote {\bibinfo {title} {Origins of scaling
  corrections in ballistic growth models},}\ }\href
  {http://dx.doi.org/10.1103/PhysRevE.90.052405} {\bibfield  {journal}
  {\bibinfo  {journal} {Phys. Rev. E}\ }\textbf {\bibinfo {volume} {90}},\
  \bibinfo {pages} {052405} (\bibinfo {year} {2014}{\natexlab{b}})}\BibitemShut
  {NoStop}%
\bibitem [{\citenamefont {Alves}\ and\ \citenamefont
  {Ferreira}(2016)}]{Alves2016}%
  \BibitemOpen
  \bibfield  {author} {\bibinfo {author} {\bibfnamefont {S.~G.}\ \bibnamefont
  {Alves}}\ and\ \bibinfo {author} {\bibfnamefont {S.~C.}\ \bibnamefont
  {Ferreira}},\ }\bibfield  {title} {\enquote {\bibinfo {title} {Scaling,
  cumulant ratios, and height distribution of ballistic deposition in $3+1$ and
  $4+1$ dimensions},}\ }\href {\doibase 10.1103/PhysRevE.93.052131} {\bibfield
  {journal} {\bibinfo  {journal} {Phys. Rev. E}\ }\textbf {\bibinfo {volume}
  {93}},\ \bibinfo {pages} {052131} (\bibinfo {year} {2016})}\BibitemShut
  {NoStop}%
\end{thebibliography}


\end{document}